\newcommand{\diracslash}[1]{#1\llap{/\kern2pt}}

\newcommand{\be}{\begin{equation}}
\newcommand{\ee}{\end{equation}}
\newcommand{\bea}{\begin{eqnarray}}
\newcommand{\eea}{\end{eqnarray}}
\newcommand{\ba}[1]{\begin{array}{#1}}
\newcommand{\ea}{\end{array}}

\newcommand{\bt}{\begin{tabular}}
\newcommand{\et}{\end{tabular}}
\newcommand{\Tr}{{\rm Tr}}

\newcommand{\ovl}{\overline}

\newcommand{\beas}{\begin{eqnarray*}}
\newcommand{\eeas}{\end{eqnarray*}}

\documentclass[preprint,prd,aps,amssymb,amsmath
,floats,nofootinbib,floatfix]{revtex4}

\DeclareSymbolFont{rsfs}{U}{rsfs}{m}{n}
\DeclareSymbolFontAlphabet{\mathrsfs}{rsfs}

\usepackage{graphicx}
\usepackage{multirow}
\usepackage{graphicx,epstopdf}

\usepackage{graphicx}
\usepackage{float}

\usepackage[utf8]{inputenc}
\usepackage[english]{babel}
\usepackage{hyperref}
\hypersetup{
    colorlinks=true,
    linkcolor=blue,
    filecolor=magenta,     
    citecolor=blue, 
    urlcolor=blue,
}
 
\usepackage{cleveref}

\begin{document}

\title{$\eta$ mesons in hot magnetized nuclear matter} 

\author{Rajesh Kumar}
\email{rajesh.sism@gmail.com}

\author{Arvind Kumar}
\email{iitd.arvind@gmail.com, kumara@nitj.ac.in}
\affiliation{Department of Physics, Dr. B R Ambedkar National Institute of Technology Jalandhar, 
 Jalandhar -- 144011,Punjab, India}
%

\def\be{\begin{equation}}
\def\ee{\end{equation}}
\def\bearr{\begin{eqnarray}}
\def\eearr{\end{eqnarray}}
\def\zbf#1{{\bf {#1}}}
\def\bfm#1{\mbox{\boldmath $#1$}}
\def\hf{\frac{1}{2}}
\def\kp{\zbf k+\frac{\zbf q}{2}}
\def\km{-\zbf k+\frac{\zbf q}{2}}
\def\hwo{\hat\omega_1}
\def\hwt{\hat\omega_2}

\begin{abstract}

		The $\eta N$ interactions are investigated in the hot magnetized asymmetric nuclear matter  using chiral SU(3) model and chiral perturbation theory (ChPT). In the chiral model, the in-medium properties of $\eta$-meson are calculated by the medium modified scalar densities under the influence of an external magnetic field. Further, in the combined approach of chiral model and ChPT,  off-shell contributions of $\eta N$ interactions  are evaluated from the ChPT effective $\eta N$ Lagrangian, and the in-medium effect of scalar densities are incorporated from the chiral SU(3) model.  We observe a significant effect of magnetic field on the in-medium mass and optical potential of $\eta$ meson. We observe a deeper mass-shift in the combined approach of  ChPT and chiral model compared to the effect of solo chiral SU(3) model. In both approaches, no additional mass-shift is  observed due to the uncharged nature of $\eta$ mesons in the presence of magnetic field.

\end{abstract}

\maketitle

\maketitle

\section{Introduction}
\label{intro}

The investigation of in-medium meson-baryon properties  under the effect of strong external magnetic field is a demanding area of research in the non-central Heavy-Ion Collisions (HICs) \cite{Cho2014,Cho2015,Gubler2016,Reddy2018,Kharzeev2008,Kharzeev2013,Fukushima2008,Skokov2009}. Besides, the presence of temperature and isospin asymmetry ,i.e., uneven numbers of neutrons and protons, lead to significant modifications in the in-medium properties of hadrons \cite{Reddy2018,Papazoglou1999,Mishra2009}. The  strong magnetic fields (of the order of $eB\sim {15{m^2_{\pi}}}$ ($5\times 10^{19}$ gauss) at large hadron collider (LHC), CERN and $eB\sim {2{m^2_{\pi}}}$ ($6.5\times 10^{18}$ gauss) at relativistic heavy ion collider (RHIC), BNL may have been produced \cite{Kharzeev2008,Fukushima2008,Skokov2009}. As the remnant move away from the collision zone, the  magnitude of the magnetic field decreases rapidly. Meanwhile, the  decaying magnetic field interacts with residual matter and as per Lenz's law, the induced current comes into the picture. These induced currents  further generate the opposite magnetic field which interacts with the primary magnetic field. This whole process slows down the decay rate of the primary magnetic field and gives it enough time to leave imprints on the mesons and hadrons \cite{Tuchin2011,Tuchin2011a,Tuchin2013,Marasinghe2011,Das2017,Reddy2018,Cho2015}. The slowing down of relaxation time  is known as chiral magnetic effect \cite{Kharzeev2013,Fukushima2008,Vilenkin1980,
	Burnier2011}. In HICs,  the time evolution of the magnetic field is still unclear, therefore to have a correct estimate of the medium's electrical conductivity and relaxation time, further study is required \cite{Reddy2018}.

In the future experiments  namely compressed baryonic matter (CBM) and antiproton annihilation at Darmstadt (PANDA)  at GSI, Germany, Japan proton accelerator research complex (J-PARC) at Japan, and  nuclotron-based
ion collider facility (NICA)  at Dubna, Russia,  we anticipate significant  research in the in-medium meson-baryons interactions  \cite{Kumar2019,Rapp2010,Kumar2020a}. On the theoretical side,  various effective models have been built to study meson-baryon interactions.  These  models are namely  quark-meson coupling (QMC) model \cite{Guichon1988,Hong2001,Tsushima1999,Sibirtsev1999,Saito1994,Panda1997},  Polyakov quark meson (PQM) model \cite{Chatterjee2012,Schaefer2010}, coupled channel approach \cite{Tolos2004,Tolos2006,Tolos2008,Hofmann2005}, chiral SU(3) model \cite{Papazoglou1999,Mishra2004a,Mishra2009,Kumar2010,
	Kumar2019}, chiral perturbation theory (ChPT) \cite{Zhong2006,Jenkins1991,Kumar2020c}, QCD sum rules  \cite{Reinders1981,Hayashigaki2000,
	Hilger2009,Reinders1985,Klingl1997,Klingl1999},  Nambu-Jona-Lasinio (NJL) model \cite{Nambu1961}, and  the Polyakov loop extended NJL  (PNJL) model \cite{Fukushima2004,Kashiwa2008,Ghosh2015}. In the present work, we have used two of the above theoretical approaches i.e. chiral SU(3) model and ChPT.

The $K/\pi/\eta-B$ interactions are much  studied in literature using various theoretical and experimental approaches \cite{Jenkins1991,Haider1986,Liu1986,Kaplan1986,Kumar2020c,Chen2017,David2018,Inoue2002}. Among these mesons, the $\eta$-meson is of special interest due to the possibility  of $\eta$-nucleon bound state formation \cite{Jenkins1991,Zhong2006,Waas1997}. The properties of $\eta$ mesons,  \cite{Peng1987,Berg1994,Chiavassa1998,Martinez1999,Averbeck2003,Agakishiev2013}, for instance, the transverse momentum spectrum near the threshold of free $N$-$N$ production  \cite{Agakishiev2013} and the $\eta$-meson production have been studied experimentally  \cite{Peng1987,Martinez1999,Agakishiev2013}.  On the theoretical side, Haider and Liu were the first to  observe that the $\eta N$ interactions
show attractive behavior and therefore, the $\eta$-meson can form  bound state with nucleons \cite{Haider1986,Liu1986}.  Chiang $et. al.$ anticipated an  optical potential $U_{\eta}$ = -34 MeV at  $\rho_0$ using the chiral coupled channel approach \cite{Chiang1991}. The authors also predicted that  the negative potential can be used to generate $\eta$-mesic atom with light/heavy nucleus. At nuclear saturation density $\rho_0$, by incorporating the leading order terms in the  coupled channel approach  a mass-shift of -20 MeV was evaluated  \cite{Waas1997}. In Ref. \cite{Wang2010}, the optical potential of -72 MeV was  anticipated.  The authors predicted the optical potential of -60 MeV at $\rho_{N}=\rho_0$ using the QMC  model \cite{Tsushima1998}.   Using ChPT and relativistic mean-field model at nuclear saturation density by including the $\eta N$  off-shell terms in the equation of motion,  the authors anticipated  optical potential of  -83 $\pm$ 5 MeV \cite{Zhong2006}.  Furthermore, using the same approach Song $et$ $al.$ obtained the negative optical potential as a function of $\eta N$ scattering length \cite{Song2008}.  The authors studied  the $\eta$ production rate and momentum dependence under the influence of isospin asymmetric HICs in Ref. \cite{Chen2017}, also the effect of $\eta N$ interactions were studied using intranuclear cascade model  under the effect of distinct medium attributes \cite{David2018}. Recently, using the combined  (chiral SU(3) model $+$ ChPT) and solo (chiral SU(3) model) approach, we derived $\eta N$ equation of motion in the non-magnetized nuclear matter and observed (-54.61) -116.83 MeV mass-shift at $\rho_0$ with $a^{\eta N}$=1.02 fm \cite{Kumar2020c}.  Evidently, the mass and optical potential of $\eta$-mesons have the model dependencies and therefore still  need more exploration.

In this article, we investigated  the magnetic field effect on the in-medium mass and optical potential of the $\eta$-meson in the hot  asymmetric nuclear matter. In this  work, we extended our previous study in the nuclear medium at zero magnetic field \cite{Kumar2020c}. First, we evaluated  the in-medium dispersion relation of $\eta$-meson using  the  $\eta N$ Lagrangian  by  the magnetically induced   scalar densities from the chiral SU(3) model \cite{Kumar2020c}.   In the second approach, we used the  scalar density of nucleons calculates using chiral SU(3) model in the dispersion relation of $\eta$-mesons which is derived from the chiral effective $\eta N$  Lagrangian of chiral perturbation theory \cite{Zhong2006}.

The chiral perturbation theory is widely used to study the   in-medium baryon-meson interactions. For the first time,  the theory was used to investigate the in-medium properties of kaons  \cite{Kaplan1986} and later it was modified by adding leading order  terms in the Lagrangian to study the interactions of $\eta$ with nucleons \cite{Jenkins1991}. The theory was also used to study astrophysical objects such as neutron stars. In the neutron star  matter, the heavy baryon ChPT was  applied to study  the kaon condensation  \cite{Brown1994,Lee1995,Kaiser1995}. Furthermore, to give correct description of $\eta N$ interactions, the next-to-leading  order terms were introduced  in the ChPT Lagrangian. Consequently, the authors anticipated more deep optical potential of $\eta$-mesons in the nuclear medium \cite{Zhong2006}.  The chiral SU(3) model is also widely used   to study the hot and dense hadronic matter \cite{Kumar2010,Zschiesche2004,Mishra2004}.  For instance, the methodology was used in the strange hadronic matter to study the in-medium properties of   kaons and antikaons  \cite{Mishra2004}. Recently, the  mass and decay width of the $\phi$ meson was also calculated in the strange hadronic matter by considering the $K \bar K$ loop at one-loop level  \cite{Kumar2020b}. The chiral SU(3) model was extended to charm SU(4) and bottom SU(5) sector to study the  properties of heavy $D$ and $B$ mesons, respectively  \cite{Mishra2004a,Mishra2009,Kumar2011}.  The chiral model is also successfully used to  anticipate the  in-medium properties of baryons and mesons in the presence of strong external magnetic field. For instance, using the combined approach of chiral  model and QCD sum rules   the  in-medium mass and decay constant of scalar, pseudoscalar, vector, and axial-vector $D$ mesons were calculated  with \cite{Kumar2020,Kumar2020a} and without incorporating the effect of the external magnetic field \cite{Kumar2014,Chhabra2017,Chhabra2017a,Chhabra2018}. Using the same combination,  the medium modified  properties of charmonia and bottomonia were  studied in the hot  magnetized asymmetric nuclear matter \cite{Kumar2019,Kumar2019a,Kumar2010}.

The outline of the present paper is as follows:
	In the coming section, we will give a brief explanation of the magnetic field effect in the present work. In   \Cref{subsec2.1.1}, we will derive the magnetic induced $\eta N$ interactions in the chiral SU(3) model whereas, in  \Cref{subsec2.1.2}, $\eta N$ formalism will be given in the  joint approach of the chiral model and chiral perturbation theory. In  \Cref{sec:3}, we will discuss the in-medium effects of  strong magnetic field on the mass of $\eta$-meson, and finally, in  \Cref{sec:4}, we will conclude our investigation.

\section{ Methodology }

\subsection{MAGNETIC FIELD INDUCED SCALAR FIELDS IN THE CHIRAL SU(3) MODEL}
\label{subsec2.1}
 The hadronic chiral SU(3) model incorporates the     trace anomaly and the non-linear realization of chiral symmetry \cite{Weinberg1968,Coleman1969,Zschiesche1997,Bardeen1969,
Kumar2020,Papazoglou1999,Kumar2019} property of the QCD. In this methodology,  the $\eta$-nucleon interactions are computed by the exchange of scalar ($\sigma$, $\zeta$, $\delta$ and $\chi$) and vector ($\omega$ and $\rho$) fields.  The glueball field $\chi$ is introduced in the model to preserve the broken scale invariance property of QCD \cite{Kumar2020}. The isospin asymmetry, $I$, of the nuclear matter is introduced  by  the addition of the scalar-isovector field $\delta$ and vector-isovector field $\rho$ \cite{Kumar2010}. In the present work, the impact of the strong magnetic field along $Z$-axis with the vector potential $A^\mu =(0,0,Bx,0)$  is studied by including the magnetic  induced Lagrangian density  to the chiral model's effective Lagrangian density \cite{Kumar2019,Reddy2018}.  Thus, we write the modified Lagrangian density of chiral model as
\be
{\cal L}_{chiral} = {\cal L}_{kin} + \sum_{ M =S,V}{\cal L}_{NM}
          + {\cal L}_{vec} + {\cal L}_0 + {\cal L}_{SB}+{\cal L}_{mag}.
\label{genlag} \ee 

Individually,
\begin{eqnarray}
{\cal L}_{NM} = - \sum_{i} \bar {\psi_i} 
\left[ m_{i}^{*} + g_{\omega i} \gamma_{0} \omega 
+ g_{\rho i} \gamma_{0} \rho \right] \psi_{i},
\label{NM}
\end{eqnarray}
\begin{eqnarray}
 {\cal L} _{vec} & = & \frac {1}{2} \left( m_{\omega}^{2} \omega^{2} 
+ m_{\rho}^{2} \rho^{2} \right) 
\frac {\chi^{2}}{\chi_{0}^{2}}
+  g_4 (\omega ^4 +6\omega^2 \rho^2+\rho^4),
\label{vec}
\end{eqnarray}
\begin{eqnarray}
{\cal L} _{0} & = & -\frac{1}{2} k_{0}\chi^{2} \left( \sigma^{2} + \zeta^{2} 
+ \delta^{2} \right) + k_{1} \left( \sigma^{2} + \zeta^{2} + \delta^{2} 
\right)^{2} \nonumber\\
&+& k_{2} \left( \frac {\sigma^{4}}{2} + \frac {\delta^{4}}{2} + 3 \sigma^{2} 
\delta^{2} + \zeta^{4} \right) 
+ k_{3}\chi\left( \sigma^{2} - \delta^{2} \right)\zeta \nonumber\\
&-& k_{4} \chi^{4} 
 -  \frac {1}{4} \chi^{4} {\rm {ln}} 
\frac{\chi^{4}}{\chi_{0}^{4}}
+ \frac {d}{3} \chi^{4} {\rm {ln}} \Bigg (\bigg( \frac {\left( \sigma^{2} 
- \delta^{2}\right) \zeta }{\sigma_{0}^{2} \zeta_{0}} \bigg) 
\bigg (\frac {\chi}{\chi_0}\bigg)^3 \Bigg ),
\label{lagscal}
\end{eqnarray}

\begin{eqnarray}
{\cal L} _{SB} =  -\left( \frac {\chi}{\chi_{0}}\right)^{2} 
\left[ m_{\pi}^{2} 
f_{\pi} \sigma
+ \big( \sqrt {2} m_{K}^{2}f_{K} - \frac {1}{\sqrt {2}} 
m_{\pi}^{2} f_{\pi} \big) \zeta \right],
\label{lsb}
\end{eqnarray}

and

\be 
{\cal L}_{mag}=-{\bar {\psi_i}}q_i 
\gamma_\mu A^\mu \psi_i
-\frac {1}{4} \kappa_i \mu_N {\bar {\psi_i}} \sigma ^{\mu \nu}F^{\mu \nu}
\psi_i
-\frac{1}{4} F^{\mu \nu} F_{\mu \nu}.
\label{lmag}
\ee

In Eq. (\ref{genlag}), the first term ${\cal L}_{kin}$   denotes the kinetic energy term, the second term ${\cal L}_{NM}$ given by Eq. (\ref{NM})  describes the nucleon-meson interaction term. In this equation, the in-medium mass of nucleons is given as $ m_{i}^{*} = -(g_{\sigma i}\sigma + g_{\zeta i}\zeta + g_{\delta i}\tau_3 \delta)$ where $\tau_3$ denotes the $z$th component of isospin quantum number and $g_{\sigma i}$, $g_{\zeta i}$ and $g_{\delta i}$ denote the coupling strengths of  scalar-isoscalar field $\sigma$,  scalar-isoscalar field $\zeta$ and scalar-isovector field $\delta$ with  nucleons ($i$=$p,n$) respectively.   The term $ {\cal L}_{vec}$ (Eq. (\ref{vec})) generates the mass of vector mesons through the interactions with scalar  mesons and quartic self-interaction terms, $ {\cal L}_{0}$ represents  the spontaneous chiral symmetry breaking where $\sigma_0$, $\zeta_0$, $\delta_0$ and $\chi_0$ symbolize the vacuum values of the $\sigma$, $\zeta$, $\delta$ and $\chi$ scalar fields, respectively.  To calculate the value of the $d$ parameter in the last term of  Eq. (\ref{lagscal}), we recall the QCD $\beta$ function at one loop level, for 
$N_{c}$ colors and $N_{f}$ flavours and is given by \cite{Schechter1980,Gomm1986}
\begin{equation}
\beta_{\rm {QCD}} \left( g \right) = -\frac{11 N_{c} g^{3}}{48 \pi^{2}} 
+ \frac{  N_{f} g^{3}}{24 \pi^{2}} +  O(g^{5}).
\label{beta}
\end{equation}

In the above expression, the first term  comes from 
the gluons self-interaction (anti-screening) and the second term comes from interactions of 
quark pairs (screening).  For  $N_c$=3 and $N_f$=3,
we estimate the value of $d$ to be 6/33, whereas for $N_c$=3 and $N_f$=2, the parameter $d$ gives the value 4/33 \cite{Schechter1980,Gomm1986,Kumar2010}.  In the present model,  
we use  $d$=0.064 \cite{Kumar2010}, which is fitted  along with the other medium parameters such as $k_i(i=1$ to $4)$ to  generate the vacuum values of  fields  ($\sigma_0$, $\zeta_0$, $\delta_0$, $\chi_0$, $\omega_0$ and $\rho_0$) and the masses of the  nucleons and $\eta$, $\eta'$ 
mesons  \cite{Papazoglou1999,Kumar2010,Kumar2019}. The values of fitted medium parameters are tabulated in \Cref{ccc}.

 Furthermore, the ${\cal L}_{SB} $ in Eq. (\ref{lsb})  denotes the explicit chiral symmetry breaking property. The  term ${\cal L}_{mag}$ accounts for the hadrons interaction with the magnetic field. 
In Eq. (\ref{lmag}), the symbol  $\psi_i$ represents a wave function  of $i$th nucleon and the second  term  describes the tensorial interaction
with the electromagnetic  tensor, $F_{\mu \nu}$. Also, the symbols  $\mu_N$ and $k_i$  represent the  nuclear magneton ($\mu_N$=$\frac{e}{2m_N}$) and anomalous magnetic moment of  $i$th nucleon, respectively.

 The non-linear coupled equations of motion  of the meson fields are obtained by solving the Euler-Lagrange equations using the total Lagrangian [Eq. (\ref{genlag})] \cite{Kumar2019,Kumar2019a} and are given as

\begin{eqnarray}
 k_{0}\chi^{2}\sigma-4k_{1}\left( \sigma^{2}+\zeta^{2}
+\delta^{2}\right)\sigma-2k_{2}\left( \sigma^{3}+3\sigma\delta^{2}\right)
-2k_{3}\chi\sigma\zeta \nonumber\\
-\frac{d}{3} \chi^{4} \bigg (\frac{2\sigma}{\sigma^{2}-\delta^{2}}\bigg )
+\left( \frac{\chi}{\chi_{0}}\right) ^{2}m_{\pi}^{2}f_{\pi}
=\sum g_{\sigma i}\rho_{i}^{s} ,
\label{sigma}
\end{eqnarray}
\begin{eqnarray}
 k_{0}\chi^{2}\zeta-4k_{1}\left( \sigma^{2}+\zeta^{2}+\delta^{2}\right)
\zeta-4k_{2}\zeta^{3}-k_{3}\chi\left( \sigma^{2}-\delta^{2}\right)\nonumber\\
-\frac{d}{3}\frac{\chi^{4}}{\zeta}+\left(\frac{\chi}{\chi_{0}} \right)
^{2}\left[ \sqrt{2}m_{K}^{2}f_{K}-\frac{1}{\sqrt{2}} m_{\pi}^{2}f_{\pi}\right]
 =\sum g_{\zeta i}\rho_{i}^{s} ,
\label{zeta}
\end{eqnarray}
\begin{eqnarray}
k_{0}\chi^{2}\delta-4k_{1}\left( \sigma^{2}+\zeta^{2}+\delta^{2}\right)
\delta-2k_{2}\left( \delta^{3}+3\sigma^{2}\delta\right) +2k_{3}\chi\delta
\zeta \nonumber\\
 +   \frac{2}{3} d \chi^4 \left( \frac{\delta}{\sigma^{2}-\delta^{2}}\right)
=\sum g_{\delta i}\tau_3\rho_{i}^{s}  ,
\label{delta}
\end{eqnarray}

\begin{eqnarray}
\left (\frac{\chi}{\chi_{0}}\right) ^{2}m_{\omega}^{2}\omega+g_{4}\left(4{\omega}^{3}+12{\rho}^2{\omega}\right) =\sum g_{\omega i}\rho_{i}^{v}  ,
\label{omega}
\end{eqnarray}

\begin{eqnarray}
\left (\frac{\chi}{\chi_{0}}\right) ^{2}m_{\rho}^{2}\rho+g_{4}\left(4{\rho}^{3}+12{\omega}^2{\rho}\right)=\sum g_{\rho i}\tau_3\rho_{i}^{v}  ,
\label{rho}
\end{eqnarray}

and

\begin{eqnarray}
k_{0}\chi \left( \sigma^{2}+\zeta^{2}+\delta^{2}\right)-k_{3}
\left( \sigma^{2}-\delta^{2}\right)\zeta + \chi^{3}\left[1
+{\rm {ln}}\left( \frac{\chi^{4}}{\chi_{0}^{4}}\right)  \right]
+(4k_{4}-d)\chi^{3}
\nonumber\\
-\frac{4}{3} d \chi^{3} {\rm {ln}} \Bigg ( \bigg (\frac{\left( \sigma^{2}
-\delta^{2}\right) \zeta}{\sigma_{0}^{2}\zeta_{0}} \bigg )
\bigg (\frac{\chi}{\chi_0}\bigg)^3 \Bigg )+
\frac{2\chi}{\chi_{0}^{2}}\left[ m_{\pi}^{2}
f_{\pi}\sigma +\left(\sqrt{2}m_{K}^{2}f_{K}-\frac{1}{\sqrt{2}}
m_{\pi}^{2}f_{\pi} \right) \zeta\right] \nonumber\\
-\frac{\chi}{{{\chi_0}^2}}(m_{\omega}^{2} \omega^2+m_{\rho}^{2}\rho^2)  = 0 ,
\label{chi}
\end{eqnarray}

respectively.

In the above equations,  the symbols  $m_\pi$, $m_K$, $f_\pi$ and $f_K$  represent  the masses and decay constants of pions and kaons, respectively. The  isospin effect is measured by the parameter  through definition, $I = -\frac{\Sigma_i \tau_{3i} \rho^{v}_{i}}{2\rho_{N}}$. Furthermore, $\rho^{s}_{i}$ and   $\rho^{v}_{i}$   describe  the  magnetic field induced scalar and vector densities of $i$th nucleons ($i=n,p$) \cite{Kumar2019,Broderick2000,Broderick2002}. Due to Landau quantization, the magnetic field interact with proton and neutron differently. For uncharged neutron the expressions for scalar and vector densities are given as

\bea
\rho^{s}_{n}&=&\frac{1}{2\pi^{2}}\sum_{s=\pm 1}\int^{\infty}_{0}
k^{n}_{\bot}dk^{n}_{\bot}\left(1-\frac{s\mu_{N}\kappa_{n}B}
{\sqrt{m^{* 2}_{n}+\left(k^{n}_{\bot}\right)^{2}}} \right)  
\int^{\infty}_{0}\, dk^{n}_{\parallel}
\frac{m^*_n}{\tilde E^{n}_{s}}\left(
f^n_{k, s}+\bar{f}^n_{k, s}\right),
\label{rhosn} 
\eea
and
\bea
\rho^{v}_{n}&=&\frac{1}{2\pi^{2}}\sum_{s=\pm 1}\int^{\infty}_{0}k^{n}_{\bot}
dk^{n}_{\bot} \int^{\infty}_{0}\, dk^{n}_{\parallel}
\left( f^n_{k, s}-\bar{f}^n_{k, s}\right), 
\label{rhovn} 
\eea

respectively. Likewise,  the scalar  and vector densities for a charged proton with the effect of Landau quantization are given by \cite{Broderick2000,Broderick2002}

\begin{equation}
\rho^{s}_{p}=\frac{|q_{p}|Bm^{*}_{p}}{2\pi^2} \Bigg [ 
\sum_{\nu=0}^{\nu_{max}^{(s=1)}}\int^{\infty}_{0}\frac{dk^p_{\parallel}}{\sqrt{(k^{p}_{\parallel})^2
+(\bar m_{p})^2}}\left( f^p_{k,\nu, s}+\bar{f}^p_{k, \nu, s}\right)
+\sum_{\nu=1}^{\nu_{max}^{(s=-1)}} \int^{\infty}_{0}\frac{dk^p_{\parallel}}{\sqrt{(k^{p}_{\parallel})^2
+(\bar m_{p})^2}}\left( f^p_{k,\nu, s}+\bar{f}^p_{k, \nu, s}\right)
\Bigg],
\label{rhosp}
\end{equation}
and

\begin{eqnarray}
\rho^{v}_{p}=\frac{|q_{p}|B}{2\pi^2} \Bigg [ 
\sum_{\nu=0}^{\nu_{max}^{(s=1)}} \int^{\infty}_{0}
dk^p_{\parallel}\left( f^p_{k,\nu, s}-\bar{f}^p_{k,\nu, s}\right)
+\sum_{\nu=1}^{\nu_{max}^{(s=-1)}} \int^{\infty}_{0}
dk^p_{\parallel}\left( f^p_{k,\nu, s}-\bar{f}^p_{k,\nu, s}\right) 
\Bigg],
\label{rhovp}
\end{eqnarray}

respectively.

In the above equations, $\bar m_{p}$=$\sqrt{m^{* 2}_{p}+2\nu |q_{p}|B}-s\mu_{N}\kappa_{p}B$ defines the effective mass of proton where symbol $\nu$ represents the  Landau levels. The effective  energy of neutron and proton  is given by

\begin{equation}
 \tilde E^{n}_{s}= \sqrt{\left(k^{n}_{\parallel}\right)^{2} +
\left(\sqrt{m^{* 2}_{n}+\left(k^{n}_{\bot}\right)^{2} }-s\mu_{N}\kappa_{n}B 
\right)^{2}}, 
\end{equation}
and
\begin{equation}
 \tilde E^{p}_{\nu, s}=\sqrt{\left(k^{p}_{\parallel}\right)^{2}+
\left(\sqrt{m^{* 2}_{p}+2\nu |q_{p}|B}-s\mu_{N}\kappa_{p}B \right)^{2}},
\end{equation}
 respectively. Also, the symbols ${f}^n_{k, \nu, s}$, $\bar{f}^n_{k, \nu, s}$,  ${f}^p_{k, s}$ and $\bar{f}^p_{k, s}$ define the  finite temperature Fermi distribution functions   for nucleon and their antinucleons, and are given as
\bea
f^n_{k, s} &=& \frac{1}{1+\exp\left[\beta(\tilde E^n_{s} 
-\mu^{*}_{n}) \right]}, \qquad
\bar{f}^n_{k, s} = \frac{1}{1+\exp\left[\beta(\tilde E^n_{s} 
+\mu^{*}_{n} )\right]}.
\label{dfn}
\eea
\bea
f^p_{k,\nu, s} &=& \frac{1}{1+\exp\left[\beta(\tilde E^p_{\nu, s} 
-\mu^{*}_{p}) \right]}, \qquad
\bar{f}^p_{k,\nu, s} = \frac{1}{1+\exp\left[\beta(\tilde E^p_{\nu, s} 
+\mu^{*}_{p} )\right]}.
\label{dfp}
\eea

\begin{table} [H]
	\centering
\caption{Values of different parameters used in the present investigation \cite{Papazoglou1999}.} \label{ccc}
	\begin{tabular}{cccccc}
		
			\hline
			\hline
			Parameter & Value & Parameter & Value &Parameter & Value\\
			\hline
			
		$k_0$ & 2.53 & $\sigma_0$ (MeV) & -93.29 & 	$g_{\sigma N}$ &10.56\\
		
		$k_1$ & 1.35 & $\zeta_0$ (MeV) & -106.8 & $g_{\zeta N }$ &-0.46\\
		
		$k_2$ & -4.77 &$\chi_0$ (MeV) & 409.8 & $g_{\delta N }$  &2.48\\
		
		$k_3$ & -2.77 & $d$ & 0.064 &	$g_{\omega N}$ &13.35\\
		
		$k_4$ & -0.218 &$g_4$ & 79.91 & $g_{\rho N}$ &5.48\\
		
		$f_K$ (MeV) & 122.14 & $\rho_0$ (${\mathrm{fm}}^{-3}$) & 0.15  & $m_{\sigma}$ (MeV) &466.5\\
		
	$m_\pi $ (MeV) & 139 & $ m_K$ (MeV) & 498 & $ f_\pi$ (MeV)  &93.29\\

	$m_{\zeta}$(MeV) & 1024.5 &$m_{\delta}$  (MeV) & 899.5 & $m_{\eta}$ (MeV)  &574.374\\
		
		$M_N$ (MeV) & 939&&&&\\
\hline
\hline
	
	\end{tabular}
	
\end{table}

\subsubsection{$\eta$N INTERACTIONS IN THE MAGNETIZED NUCLEAR MATTER}
\label{subsec2.1.1}

In this subsection, we evaluate the in-medium mass of  $\eta$ mesons via dispersion relation  
in hot magnetized asymmetric  nuclear matter. The medium modified $\eta$ meson mass is obtained in terms of scalar and vector fields of the chiral model which are solved by considering the interactions of nucleons with $\eta$ mesons in the presence of an external magnetic field. These scalar and vector fields modify the   scalar and vector densities of the nucleons which in result modifies the self-energy of the $\eta$ mesons.

The  $\eta N$ interaction Lagrangian is given as

	\begin{equation}
		\label{etasN}
		{\mathcal L}_{{\eta N}} = {\mathcal L}_{{RT}}+ {\cal L}_{\eta SB} + 	{\cal L }_{d_1}^{BM}+	{\cal L }_{d_2}^{BM}. 
	    \end{equation}
	    The individual terms are given in detail as follows

	\begin{enumerate}

		    \item ${\mathcal L}_{{RT}}$, The	first range term:
		    
		    The first term in the $\eta N$ Lagrangian comes from the first range term 

\begin{equation}
\label{pikin}
 {\mathcal L}_{{\mathrm{1st range term}}} =  Tr (u_{\mu} X u^{\mu}X +X u_{\mu} u^{\mu} X) ,
\end{equation}
where  $u_{\mu} =-\frac{i}{2} \left[u^{\dagger}(\partial_{\mu}u) 
-u (\partial_{\mu}u^\dagger) \right]$ and $
u$=$ \text{exp}\left[ \frac{i}{\sqrt{2}\sigma_{0}}P\gamma_{5}\right]$. In the present investigation, we have taken the interactions up to second order. The  $X$ and $P$, represent the scalar and pseudoscalar   meson matrices  \cite{Zhong2006}, respectively and are explicitly given as
\be
\label{smatrix}
X=\frac{1}{\sqrt{2}}\sigma^a \lambda_a=
\left( \begin{array}{ccc}
   (\delta  +\sigma)/\sqrt{2} & \delta^+ & \kappa^+\\   
    \delta^- & (-\delta+\sigma)/\sqrt{2} & \kappa^0 \\
   \kappa^- & \ovl{\kappa^0}& \zeta 
\end{array} \right),
\ee

and 

\be
P=\frac{1}{\sqrt{2}}\pi_a \lambda^a
=\left (\begin {array}{ccc} 
  \frac{1}{\sqrt{2}}\left ( \pi^0+{\frac {\eta}{\sqrt {1+2\,{w}^{2}}}}\right )&\pi^{+}
 &2\,{\frac {K^{+}}{w+1}}\\
  \noalign{\medskip}\pi^{-}&\frac{1}{\sqrt{2}}\left 
 (- \pi^0+
 {\frac {\eta}{\sqrt {1+2\,{w}^{2}}}}\right )&2\,{\frac { K^0
  }{w+1}}\\\noalign{\medskip}2\,{\frac {K^-}{w+1}}&2\,{\frac { 
 \ovl{K}^0}{w+1}}&-{\frac {\eta\,\sqrt {2}}{\sqrt {1+2\,{w}^{2}}}}
 \end {array}\right ).
 \label{psmat}
 \ee
 
In Eq. (\ref{pikin}), the calculations of the axial current of pions and kaons result in the following relations

\be
\label{zeta0}
\sigma_0 = -f_{\pi} \qquad \zeta_0 = -\frac{1}{\sqrt{2}}(2 f_K - f_{\pi}) ,
\ee 
for the vacuum values of the scalar condensates $\sigma$ and $\zeta$ found in the linear $\sigma$-model 
\cite{Papazoglou1999}.  In the Eq. (\ref{psmat}), the re-normalization factor 
$w=\sqrt{2}\zeta_0/\sigma_0$ is incorporated 
to obtain  the canonical form 
of the kinetic energy terms \cite{Papazoglou1999}.  The matrix $P$
reduces to the matrix normally used in 
in chiral perturbation theory \cite{Zhong2006} for  $w=1$ (i.e. $f_{\pi}$=$f_K$). The advantage of  $w\neq 1$
is that the  SU(3)$_V$ breaking effects  are accounted in the $P$ matrix for even at lowest order \cite{Zhong2006}.

	\item ${\cal L}_{\eta SB}$, The	mass term:
	
	The second term in  Eq. (\ref{etasN}), represents the scale breaking term of the chiral model Lagrangian, which is given by

 \begin{equation}
\label{esb-gl}
 {\cal L}_{SB}  =  
     -\frac{1}{2} \Tr A_p \left(uXu+u^{\dagger}Xu^{\dagger}\right),
\end{equation}

with $A_p$ as a diagonal matrix, given as

\begin{equation}
A_p=\frac{1}{\sqrt{2}}
\left( \begin{array}{ccc}
   m_{\pi}^2 f_{\pi}& 0& 0\\   
    0 & m_\pi^2 f_\pi& 0\\
   0 & 0& 2 m_K^2 f_K
-m_{\pi}^2 f_\pi
\end{array} \right).
\end{equation}

 The  $\eta$ meson vacuum mass  is extracted from the  Lagrangian [Eq. (\ref{esb-gl})] and given as

	\begin{equation}
		m_{\eta}=\frac{1}{f}\sqrt{\left(3 m_\pi^2  f_K m_K^2+\frac{8 f_K^2 m_K^2}{f_\pi^2} -\frac{4f_K  m_\pi^2}{f_\pi} \right)}.
	\end{equation}
	
	Using the values of various constants, the value of  $m_{\eta}$ turns out to be 574.374 MeV which is with an accuracy of 4.9 $\%$  of experimental mass, i.e., 547.862 MeV \cite{PDG2020}. Moreover,   using Gell-Mann Okubo mass formula under octet approximation, the authors calculated the vacuum mass of $\eta$-meson as 567 MeV which  is with an accuracy of 3.6 $\%$ of physical mass \cite{Burakovsky1997}. It has been observed that the vacuum mass of $\eta$-meson has  model dependencies \cite{Burakovsky1997} but here in the current scenario, the  in-medium mass-shift of $\eta$-meson  is nearly the same for both  obtained masses and therefore can be neglected.

	\item ${\cal L }_{d_1}^{BM}+	{\cal L }_{d_2}^{BM}$, The	 $d$  terms:

 The last term in the Eq. (\ref{etasN})
is called another range term which basically arises from the baryon-meson interaction Lagrangian terms of chiral model  \cite{Mishra2004a,Mishra2006} and are given as
\begin{equation}
{\cal L }_{d_1}^{BM} =\frac {d_1}{2} Tr (u_\mu u ^\mu)Tr( \bar B B),
\end{equation}
and,
\begin{equation}
{\cal L }_{d_2}^{BM} =d_2 Tr (\bar B u_\mu u ^\mu B).
\end{equation}
In above,  $B$ denotes the baryon matrix, given as

\be
\label{bmatrix}
B=\frac{1}{\sqrt{2}}b^a \lambda_a=
\left( \begin{array}{ccc}
   \frac{\Sigma^0}{\sqrt{2}} +\frac{\Lambda^0}{\sqrt{6}}& \Sigma^+ & p\\   
    \Sigma^- & -\frac{\Sigma^0}{\sqrt{2}} +\frac{\Lambda^0}{\sqrt{6}} & n \\
   \Xi^- & \Xi^0& -2 \frac{\Lambda^0}{\sqrt{6}}
\end{array} \right).
\ee

\end{enumerate}
	
The explicit form of above three terms are inserted in Eq. (\ref{etasN}) and the interaction Lagrangian is given as 
\begin{eqnarray} \label{etaN}
\mathcal{L_{\eta }}  &=&
 \left( \frac{1}{2}-\frac{\sigma ^\prime + 4 \zeta ^\prime (2 f_K-f_\pi) }{\sqrt{2}f^2} \right) \partial^{\mu}\eta\partial_{\mu}\eta \nonumber\\
 &-&\frac{1}{2}\left(
  m_{\eta}^2
-\frac{(\sqrt{2}\sigma ^\prime -4 \zeta ^\prime )m^2_\pi f_\pi + 8 \zeta ^\prime m^2_K f_K}{\sqrt{2} f^2}
 \right) \eta^2\nonumber\\
 &&+\frac{d'}{f^2} \left( \frac{\rho^s_p+\rho^s_n}{4}  \right) \partial^{\mu}\eta\partial_{\mu}\eta.
\end{eqnarray}

 In above,  the  fields $\sigma'(=\sigma-\sigma _0)$,
$\zeta'(=\zeta-\zeta_0)$ and  $\delta'(=\delta-\delta_0)$ are the  digression of the expectation values  of scalar fields
from their vacuum expectation values, the constant $f$, is equal to $\sqrt{f_\pi^2+2(2 f_K - f_\pi)^2}$ and  the parameter  $d'$=$3d_1+d_2$.

At the mean-field level, 
the equation of motion for the $\eta$ meson field is simplified to
  \begin{eqnarray} 
&& \partial^{\mu}\partial_{\mu} \eta +\left(
  m_{\eta}^2-\frac{(\sqrt{2}\sigma ^\prime -4 \zeta ^\prime )m^2_\pi f_\pi + 8 \zeta ^\prime m^2_k f_k}{\sqrt{2} f^2}
 \right)\eta  \nonumber\\
&&+\frac{2d'}{f^2} \left( \frac{\rho^s_p+\rho^s_n}{4} -\frac{\sigma ^\prime + 4 \zeta ^\prime (2 f_K-f_\pi) }{\sqrt{2}} \right) \partial^{\mu} \partial_{\mu}  \eta=0.
\end{eqnarray}

Furthermore, the dispersion relation for $\eta$ meson field is obtained  by Fourier transformation of the above equation
\begin{equation}
-\omega^2+ {\vec k}^2 + m_\eta^2 -\Pi^*(\omega, |\vec k|)=0,
\label{drk}
\end{equation}

where $\Pi^*$ symbolize the in-medium self-energy of  $\eta$ meson, and it is explicitly given as
\begin{eqnarray}
\Pi^* (\omega, |\vec k|) &= & 
  \frac{(\sqrt{2}\sigma ^\prime -4 \zeta ^\prime )m^2_\pi f_\pi + 8 \zeta ^\prime m^2_K f_K}{\sqrt{2} f^2}
  +\frac{2d'}{f^2} \left( \frac{\rho^s_p+\rho^s_n}{4} \right)
(\omega ^2 - {\vec k}^2) \nonumber\\
&-& \frac{2}{f^2} \left( \frac{\sigma ^\prime + 4 \zeta ^\prime (2 f_K-f_\pi) }{\sqrt{2}} \right)
(\omega ^2 - {\vec k}^2).
\label{sek}
\end{eqnarray}

In the asymmetric nuclear  matter, the in-medium mass of $\eta$ meson is evaluated by solving  Eq. (\ref{drk}) under the condition,
$m_{\eta}^*=\omega(|\vec k|$=0). The parameter $d'$ in the  expression of self energies
is estimated from the empirical value of scattering length $a^{\eta N}$ of $\eta$ meson \cite{Zhong2006}, whose expression is given as


\begin{eqnarray}
d' &=& \frac{ f^2}{2 \pi \left (1+\frac{m_\eta}{M_N}\right )} \frac{ a^{\eta N}}{m^2_\eta} +\frac{\sqrt{2}g_{\sigma N}}{m^2_\sigma}-\frac{4 \sqrt{2} (2f_K-f_\pi) g_{\zeta N}}{m^2_\zeta}  \nonumber \\
&-& \left( \frac{\sqrt{2} g_{\sigma N}}{m^2_\sigma}-\frac{4 g_{\zeta N}}{m^2_\zeta} \right )\frac {m^2_\pi f_\pi} {\sqrt{2} m^2_\eta}-\tau _3 \frac{4\sqrt{2} g_{\delta N}m^2_K f_K}{m^2_\delta m^2_\eta},
\label{dek}
\end{eqnarray}

where $m_{\sigma}$, $m_{\zeta}$
, $m_{\delta}$
 and $m_{N}$ denote the vacuum masses of  the fields $\sigma$, $\zeta$, $\delta$ and nucleons, respectively and their values are given in \Cref{ccc}. Using, the in-medium mass of $\eta$ mesons, the optical
potential for $\eta$-meson for finite momentum  \cite{Mishra2008,Mishra2009} in the nuclear matter is given by 
\begin{equation}
U^*_{\eta}(\omega, \textbf{k}) = \omega (\textbf{k}) -\sqrt {\textbf{k}^2 + m^2_{\eta}},
\label{opk}
\end{equation}
and for zero momentum, the relation becomes
\begin{equation}
U^*_{\eta} =\Delta m_\eta^*={m_{\eta}^*}-m_{\eta}.
\end{equation}

\subsubsection{\label{subsec2.1.2} FUSION OF CHIRAL PERTURBATION THEORY  AND CHIRAL SU(3) MODEL}

Chiral Perturbation theory (ChPT) is one of the phenomenological approach to study the low-energy dynamics of QCD with an effective field theory  Lagrangian  based on  the underlying  chiral symmetry of quantum chromodynamics \cite{Zhong2006}. In this,  the SU(3)$_{\mathrm{L}}\times$SU(3)$_{\mathrm{R}}$ Lagrangian describing the pseudoscalar
mesons  and baryons interactions is given as

\begin{eqnarray}\label{LL}
{\mathcal{L}_{\text{ChPT}} }={\mathcal{L}_{P} }+{\mathcal{L}_{P
B} },
\end{eqnarray}
where the pseudoscalar mesonic term, $\mathcal{L}_{P}$ is taken  up to second chiral
order \cite{Zhong2006,Kaplan1986} and is given by \cite{Zhong2006},
\begin{eqnarray}
{\mathcal{L}_{P} }&=&\frac{1}{4}f_\pi^{2}\textrm{Tr}
 \partial^{\mu}\Sigma\partial_{\mu}\Sigma^{\dagger}
 +\frac{1}{2}f_\pi^2 B_0
  \left\{\mbox{Tr} M_{q}(\Sigma-1)+\mathrm{H.c.}\right\}.
\end{eqnarray}
In above equation, $\Sigma=\xi^2=\exp{(i\sqrt{2}P/f_\pi)}$, the symbol $B_0$ represent the connection with order
parameter of spontaneously broken chiral symmetry and  $M_{q}=\mbox{diag}\{m_{q}, m_{q}, m_{s}\}$
being the current quark mass matrix. The second term in the  Eq. (\ref{LL}),
$\mathcal{L}_{P B}$  defines the leading order and next-to leading order baryon-meson interactions \cite{Kaplan1986}. The off-shell  terms are developed by using 
heavy baryon chiral theory \cite{Jenkins1991}.
However, the former theory has additional properties such as
 quantum corrections and Lorentz invariance. The  properties of the nuclear system has been described successfully by using the off-shell Lagrangian  and the  higher-order terms of this next-to-leading order Lagrangian are also studied \cite{Park1993}.
In the present article,
we have limited our calculations up to the  small
momentum scale, $Q^{2}$ without loop contributions (for s-wave
$\eta {N}$ scattering) because the  higher order corrections are
suppressed \cite{Zhong2006}.

 By using the
heavy-baryon approximation and expanding the Eq. (\ref{LL})  up to the order of $1/f_\pi^2$, we get the $\eta N$ Lagrangian as
\begin{eqnarray} \label{teffL}
\mathcal{L_{\eta N }}  &=&
 \frac{1}{2}\partial^{\mu}\eta\partial_{\mu}\eta
 -\frac{1}{2}\left(
  m{^\prime}_{\eta}^2
-\frac{\Sigma_{\eta
\mathrm{N}}}{f_\pi^2}\bar{\Psi}_{\mathrm{N}}\Psi_{\mathrm{N}}
 \right) \eta^2+\frac{1}{2}\frac{\kappa}{f_\pi^2}\bar{\Psi}_{\mathrm{N}}\Psi_{\mathrm{N}}\partial^{\mu}\eta\partial_{\mu}\eta.
\end{eqnarray}
In above equation,  $m_{\eta}$ represent  the mass  of $\eta$-meson calculated in ChPT and is evaluated by relation $m{^\prime_{\eta}}^2=\frac{2}{3}B_0 (m_q+2m_s)$. In this mass relation,  $m_{q(s)}$ defines the mass of light (strange) quarks \cite{Burakovsky1997}. We have used the same value of  $\eta$ meson vacuum mass i.e. $m{^\prime}_{\eta}$=$m_{\eta}$= 574.374 MeV in the ChPT+chiral model calculations  for consistency with the chiral SU(3) model. Also, the
$\Sigma_{\eta\mathrm{N}}$, the $\eta N$ sigma term  and the $\kappa$ term is determined by relations
\begin{eqnarray} \label{sigmaexp}
\Sigma_{\eta\mathrm{N}}
=-\frac{2}{3}[a_{1}m_{q}+4a_{2}m_{s}+2a_{3}(m_{q}+2m_{s})],
\end{eqnarray}

and 
\begin{eqnarray} \label{akap}
\kappa =4\pi f_\pi^2
 \left(
  \frac{1}{m_{\eta}^2}+\frac{1}{m{^\prime}_{\eta}M_{\mathrm{N}}}
 \right)
a^{\eta\mathrm{N}} -\frac{\Sigma_{\eta\mathrm{N}}}{m_{\eta}^2},
\end{eqnarray}

respectively.  The $a$ terms in the Eq. (\ref{sigmaexp})  corresponds to
the chiral breaking  effects and  are fitted from the parameter
$\Sigma_{\mathrm{KN}}=380\pm 100$ MeV, where $\pm 100$
MeV reflects the uncertainty \cite{Lyubovitskij2001,Dong1996,Hatsuda1994,
Brown1994,Georgi1984,Politzer1991,Lee1995,Zhong2006}. The parameter, $\kappa$ is
estimated from the $\eta$N scattering length \cite{Zhong2006} with the range of  
$a^{\eta\mathrm{N}}$  values i.e. 0.91 $\sim$ 1.14 fm, which is assumed from the empirical investigations \cite{Green2005,Renard2002,Arndt2005,Green1999,Zhong2006}.

The  equations of motion for $\eta N$ interactions in the unified approach of chiral SU(3) model and ChPT can be written as
by
\begin{eqnarray}
 \left(
  \partial_{\mu}\partial^{\mu}
  +m_{\eta}^2
  -\frac{\Sigma_{\eta N}}{2f_\pi^2}\langle
\bar{\Psi}_{\mathrm{N}}\Psi_{\mathrm{N}} \rangle
  +\frac{\kappa}{2f_\pi^2}\langle
\bar{\Psi}_{\mathrm{N}}\Psi_{\mathrm{N}} \rangle\partial_{\mu}\partial^{\mu}
 \right)
 \eta
 = 0,
 \label{eqm}
\end{eqnarray}

where $ \langle
\bar{\Psi}_{\mathrm{N}}\Psi_{\mathrm{N}}\rangle \equiv \rho^s_{N}$=$\left(\rho^s_{p}+\rho^s_{n} \right)$  is the magnetic field influenced scalar
density of nucleon calculated within the chiral SU(3) model. The plane wave decomposition of Eq. (\ref{eqm}) gives
\begin{eqnarray} \label{decom}
-\omega^2+\vec {\textbf{k}}^2+m_{\eta}^2
-\frac{\Sigma_{\eta\mathrm{N}}}{2f_\pi^2}\rho^s_{N}
+\frac{\kappa}{2f_\pi^2}\rho^s_{N}
 \left(-\omega^2+\vec{\textbf{k}}^2\right)=0.
\end{eqnarray}
By solving the above quadratic equation, we get
\begin{equation} \label{effmetadef}
\omega=\sqrt{{m_{\eta}^*}^2 +\vec{\textbf{k}}^2},
\end{equation}
and the explicit  expression of magnetic field induced mass of $\eta$ meson, $m_{\eta}^*$ is given by
\begin{eqnarray}
m_{\eta}^*
=\sqrt{
  \left(m_{\eta}^2-\frac{\Sigma_{\eta\mathrm{N}}}{2f_\pi^2}\rho^s_{N}\right)
  \Big/ \left(1+\frac{\kappa}{2f_\pi^2}\rho^s_{N}\right)
      }.
      \label{metac}
\end{eqnarray}

The last two terms  of the Eq.\
(\ref{decom}) gives the $\eta$-meson self-energy
\begin{eqnarray}
\Pi^*(\omega,\vec{\textbf{k}})
=\Big (-\frac{\Sigma_{\eta\mathrm{N}}}{2f_\pi^2}
 +\frac{\kappa}{2f_\pi^2}
  (-\omega^2+\vec{\textbf{k}}^2)\Big ) \rho^s_{N},
  \label{sekchpt}
\end{eqnarray}
where  $\omega$ is  $\eta$-meson single-particle energy
 and  $\vec{\textbf{k}}$ is the momentum.


\section{Results and Discussions}
\label{sec:3}

	In this section,  we discuss  the magnetic field induced optical potential of $\eta$ meson  evaluated using two approaches i.e. (i) chiral SU(3) model in   \Cref{subsec:3.1}    and (ii) ChPT + chiral SU(3) model  in   \Cref{subsec:3.2}. In  both methodologies, we have taken the values of  scattering length, $a^{\eta_N}$ in the range 0.91-1.14 fm.  We start by discussing the in-medium behavior of nucleon scalar densities under the influence of a strong magnetic field for different values of nuclear density, isospin asymmetry, and temperature.

\begin{figure}[h]
\includegraphics[width=16cm,height=16cm]{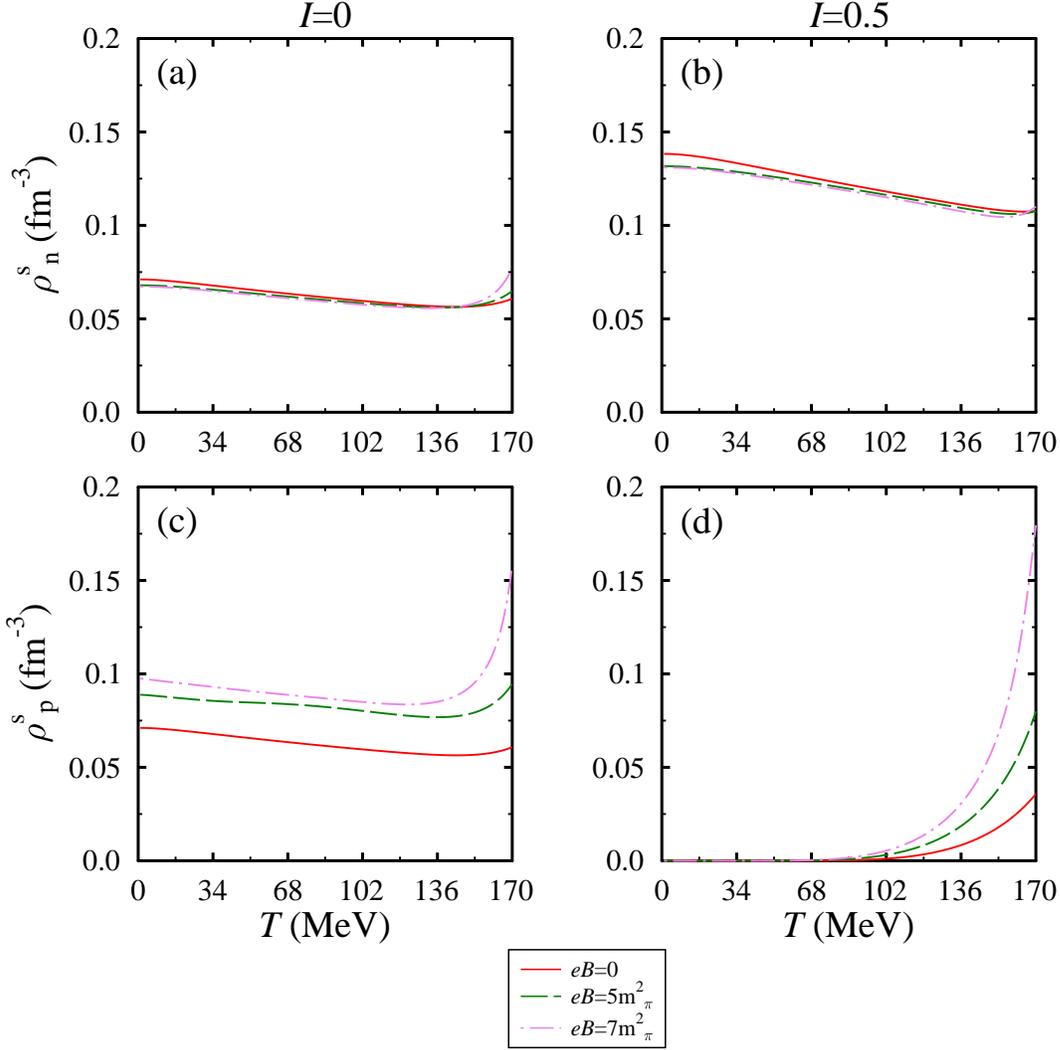}
\caption{(Color online) The in-medium scalar density of nucleons at $\rho_0$. }
\label{f1}
\end{figure}

	In \Cref{f1}, at nuclear saturation density, we illustrate the scalar density of neutron and proton as a function of temperature for zero and non-zero values of the magnetic field. In the left (right) column of the figure, we present the scalar densities for symmetric (anti-symmetric) nuclear matter. For symmetric nuclear matter and zero magnetic field, we observe the same behavior of neutron and proton scalar density with temperature. The scalar densities slowly decrease linearly up to $T \approx$150 MeV and start increasing for higher values of temperature.  These modifications  reflect the interplay between the contributions from higher momenta states and the thermal distribution functions in the scalar density expressions [see Eqs. (\ref{rhosp}) and (\ref{rhosn})]. Further, on increasing the magnetic field the proton and neutron scalar density behave unevenly, for a particular value of temperature, the proton scalar density increases significantly whereas the neutron scalar density slightly decreases. The additional effects in proton scalar density are because of the charged nature of proton, the positively charged proton interacts with the magnetic field and experiences  Landau quantization and contributions from the anomalous magnetic moment [see Eq. (\ref{rhosp})]. Whereas due to uncharged nature, the neutron only experiences contributions from the anomalous magnetic moment [see Eq. (\ref{rhosn})].  Moreover, the transition temperature decrease with the increase in the magnetic field which supports the restoration of chiral symmetry \cite{Kumar2020}. On the other side, for asymmetric nuclear matter [sub-plot (b) and (d)], even for zero value of magnetic field we find unequal values of proton and nucleon scalar density. 	In  the  chiral SU(3) model, the scalar densities of protons and neutrons are calculated through Eq. (\ref{rhosp}) and Eq. (\ref{rhosn}), respectively and these equations include the effect of in-medium scalar and vector fields \cite{Papazoglou1999}. In asymmetric nuclear matter the iso-scalar(vector) $\delta(\rho)$ field show non-zero contributions which eventually leads to unequal values of proton  and neutron densities \cite{Kumar2020c}. The neutron scalar density modifies significantly in asymmetric nuclear medium and decreases with the increase in the temperature whereas the proton scalar density shows zero value up to $T  \approx$ 90 MeV and then increases rapidly. Naively, the value of proton scalar density should be zero for $I$=0.5 but at higher temperatures, despite $\rho^v_p$=0, the proton condensate ($\bar p p$) still  populates in the nuclear matter. The inclusion of magnetic field does significant changes in the proton scalar density whereas the neutron scalar density shows a small decrement with the increasing magnetic field.

\begin{figure}[h]
\includegraphics[width=16cm,height=16cm]{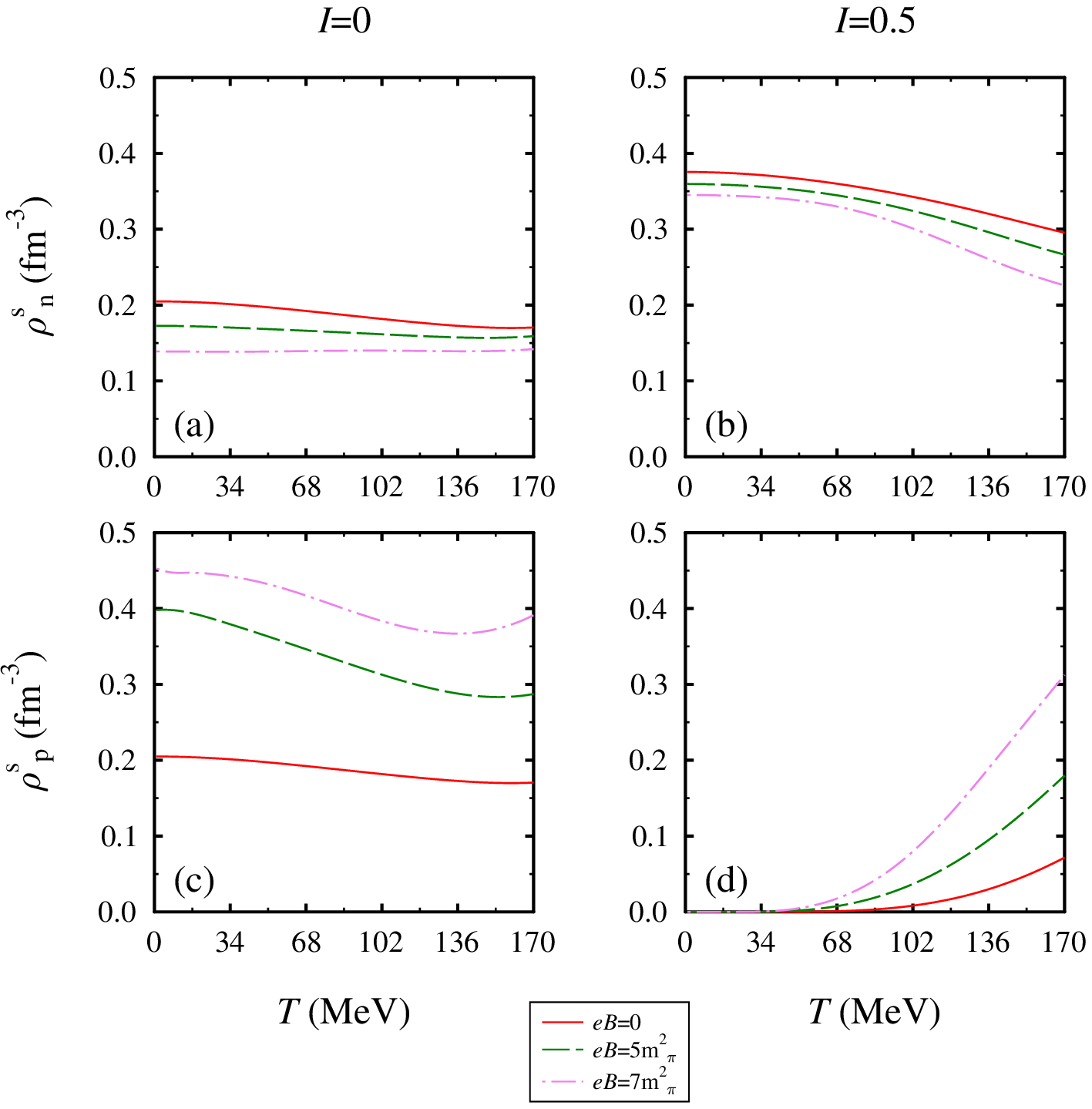}
\caption{(Color online) The in-medium scalar density of nucleons at 4$\rho_0$. }
\label{f2}
\end{figure}

	In \Cref{f2}, we plot the scalar densities for the same values of medium parameters but $\rho_N$=4$\rho_0$. On the same line, at $eB$=0 and $I$=0, we observe similar behavior of proton and neutron scalar density. When we move from zero to non-zero values of magnetic field strength, for a particular value of temperature we observe the proton scalar density increase appreciably whereas the neutron scalar density slightly decreases. This is again due to the additional energy levels in the protons due to the magnetic field intervention. It is to be noted that the effect of the magnetic field is more pronounced in the high density regime. Furthermore, in the highest value of medium asymmetry, as a function of temperature, we observe that the proton scalar density remains zero up to $T \approx$50 MeV and further increases non-linearly with the increase in the magnetic field. On the contrary, the neutron scalar density modifies appreciably in the asymmetric matter, it decreases with the increase in temperature and magnetic field.  In the symmetric nuclear matter, the observed scalar densities  at zero magnetic field are in agreement with the results of the relativistic mean-field model \cite{Zhong2006,Song2008}.

\subsection{Impact of Magnetic Field on the $\eta$ Mesons in Chiral Model}
	\label{subsec:3.1}

\begin{figure}[h]
\includegraphics[width=16cm,height=21cm]{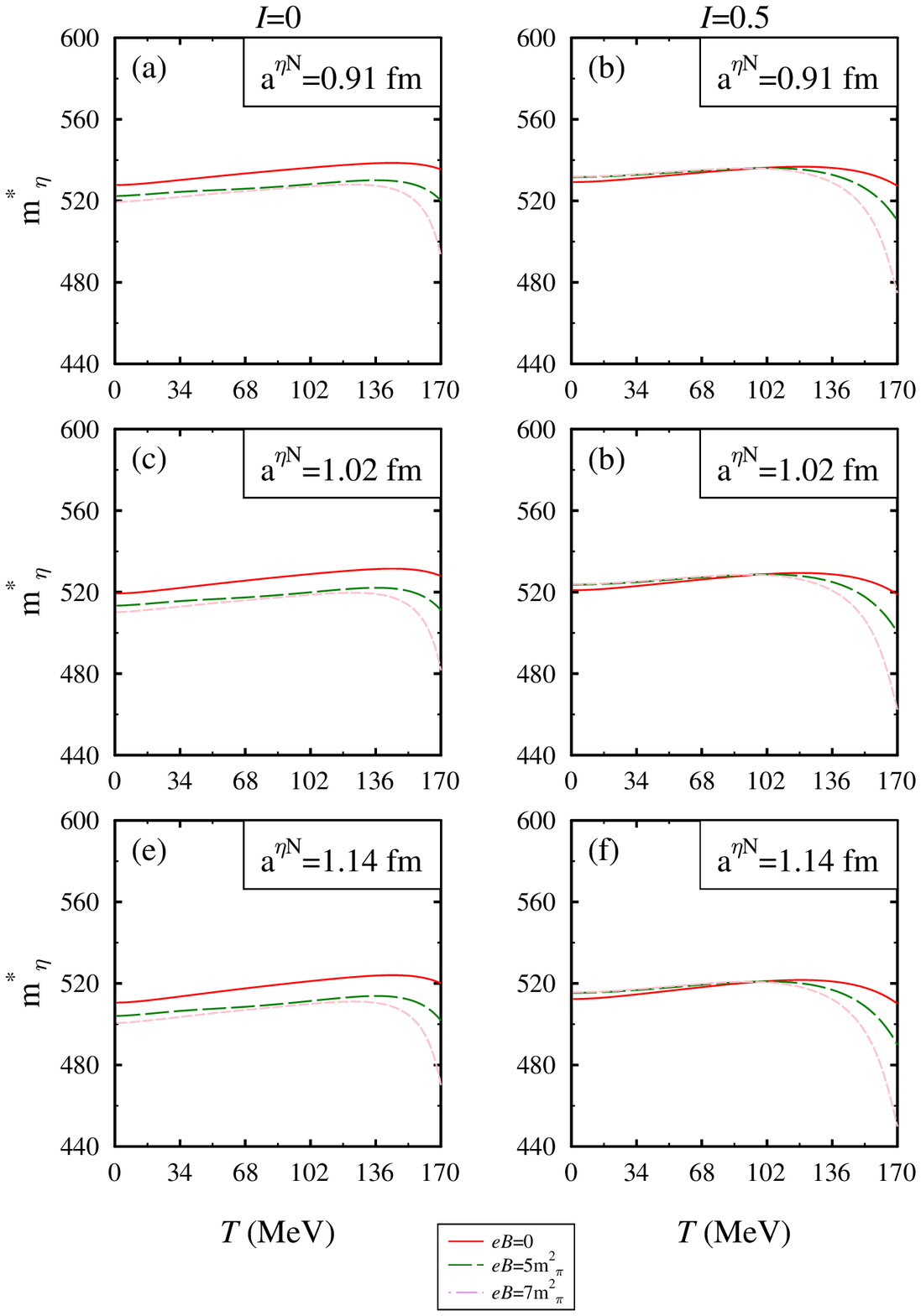}
\caption{(Color online) The in-medium $\eta$ meson mass in chiral model at $\rho_0$ . }
\label{f3}
\end{figure}

\begin{figure}[h]
\includegraphics[width=16cm,height=21cm]{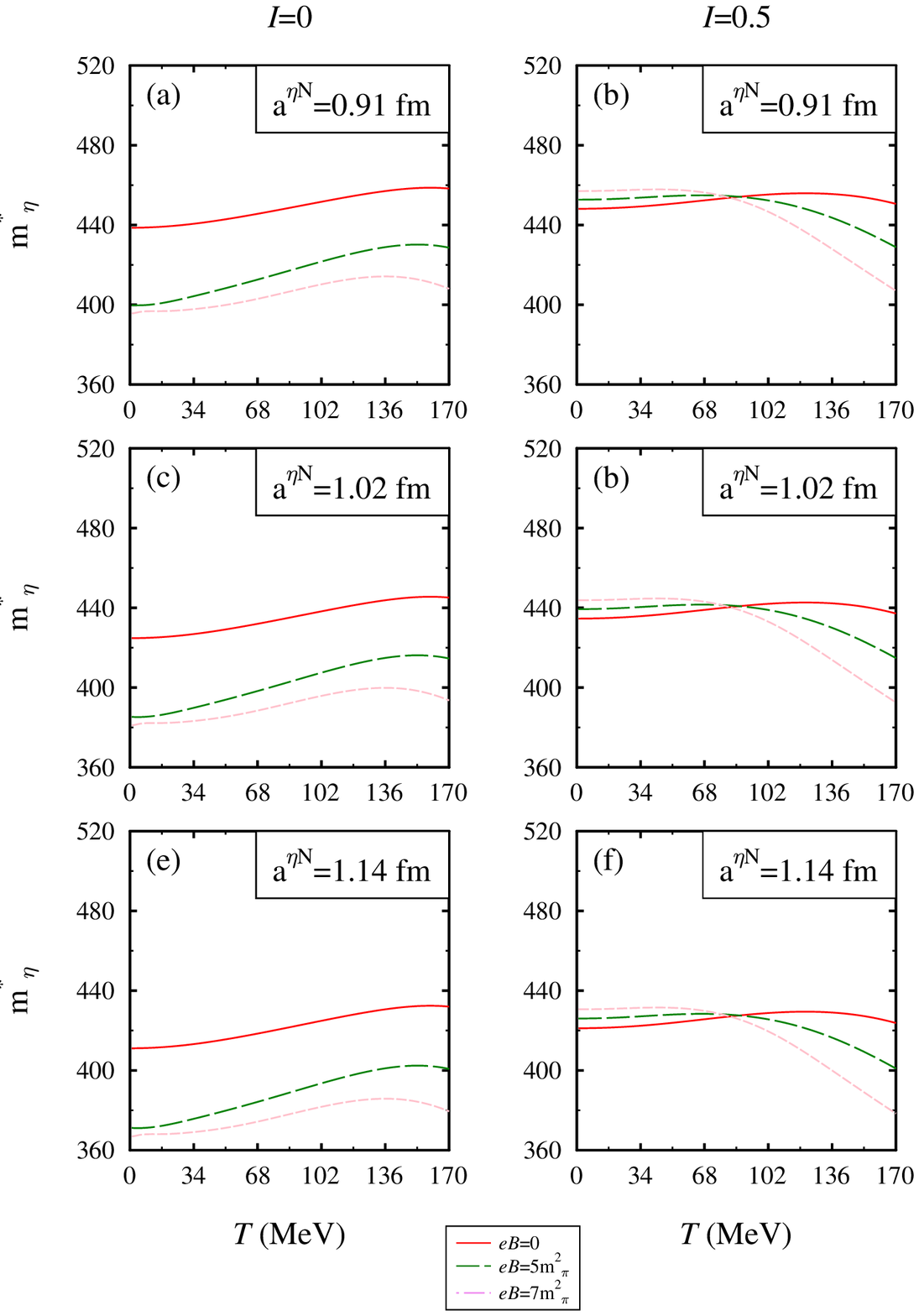}
\caption{(Color online) The in-medium $\eta$ meson mass in chiral model at 4$\rho_0$ . }
\label{f4}
\end{figure}

	In this section, we present the result and discussion for the in-medium mass of $\eta$-meson calculated in the chiral SU(3) model under the influence of the external magnetic field. In \Cref{f3,f4}, we illustrate the in-medium $\eta$ mass as a function of temperature for different other  parameters such as isospin asymmetry, magnetic field, and  scattering length  at $\rho_N$=$\rho_0$ and 4$\rho_0$, respectively. In \Cref{f3}, for any value of $I$, $eB$ and $a^{\eta N}$, we observe the in-medium mass gradually increases with the increase in temperature up to a certain value of temperature and then it starts decreasing.  This behavior represents the opposite reflection of in-medium scalar densities plotted in \Cref{f1} as the  expression of $\eta$-meson  [see Eq. (\ref{metac})] has an inverse dependence on the sum of scalar densities of nucleons. In symmetric nuclear matter, the impact of the magnetic field leads to a more attractive contribution in the in-medium $\eta$ mass for a particular value of temperature, scattering length. With the increase in the magnetic field, we observe the transition point (i.e. the temperature where in-medium mass starts decreasing as a function of temperature) moves towards the lower temperature side. As discussed earlier, the medium modified mass of $\eta$-meson has indirect dependence on the sum of the  nucleon's  scalar densities and therefore it shows opposite behavior to the scalar densities. However, in the asymmetric nuclear  matter, we observe that the in-medium mass slowly increases for zero value of the magnetic field as was observed for the symmetric matter. This is because the  in-medium mass depends upon the sum of scalar densities with no additional parameter dependence. Further, at $I \neq$0, we observe a  little change in  $\eta$ mass for the lower temperature region whereas a substantial change in the higher temperature region concerning the magnetic field. This  is because in the highly asymmetric matter for lower (higher) temperatures, we have a negligible (substantial) contribution to the proton scalar density. It is to be noted that due to the uncharged nature of $\eta$-meson   it does not couple with magnetic field directly and therefore does not show any additional contributions from Landau energy levels as was observed for charged $D$ mesons \cite{Kumar2020,Reddy2018,Kumar2020a}. From \cref{f3}, we also anticipate the effect of scattering length. When we change $a^{\eta N}$  from 0.91 to 1.14 fm, we observe a significant  decrement in the effective mass for a particular value of  magnetic field, and temperature. This is because  of the parameter $d^\prime$'s direct relationship  with the scattering length in Eq. (\ref{dek}). The  $d^\prime$ parameter gives attractive contribution to the $\eta$ in-medium mass through equation of motion  [Eq. (\ref{drk})] and self-energy [Eq. (\ref{sek})].

	In \Cref{f4}, we plot the in-medium mass of $\eta$-meson for same values of medium parameters but $\rho_N$=4$\rho_0$. In the left panel, at $eB$=0 and $I$=0, we observe a similar trend of in-medium mass as a function of temperature as was observed for $\rho$=$\rho_0$. For high nuclear density, we observe a significant drop in the mass of $\eta$-meson. The drop in effective mass increase with the increase in the magnetic field and scattering length. When we move from symmetric nuclear matter to asymmetric nuclear matter, we observe the cross-over behavior of in-medium mass as a function of temperature for a particular value of scattering length. This is due to a similar reason that was discussed for the nuclear saturation density case. The difference is, here in the high density regime the proton scalar density populates little (but greater than the $\rho$=$\rho_0$ case)  in the lower temperature regime and substantially in the higher temperature regime. Also, the effect of the magnetic field is more pronounced in the high density regime.

\begin{figure}[h]
\includegraphics[width=16cm,height=16cm]{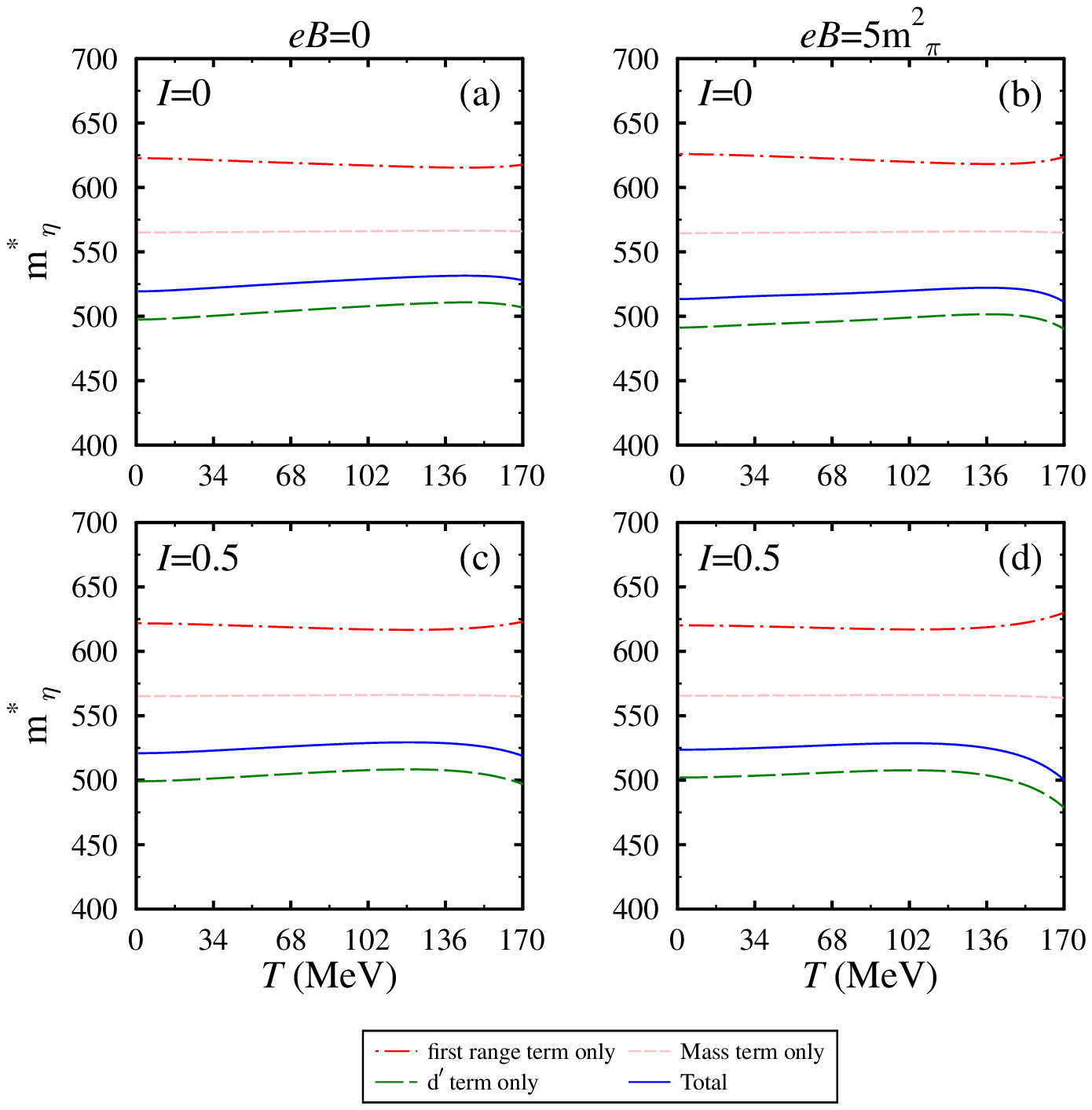}
\caption{(Color online) The different terms of  in-medium $\eta$ meson mass in chiral model at $\rho_0$ and $a^{\eta N}$=1.02 fm. }
\label{f5}
\end{figure}

Furthermore, for a better understanding of in-medium $\eta$-meson mass, in \Cref{f5}, we plot the individual terms of $\eta$-meson's self-energy. The  expression of  self-energy  [Eq. (\ref{sek})] have three interaction terms: (i) first range term (ii) mass term and (iii) $d^\prime$ term. At nuclear saturation density, in this figure, we show the contributions of the individual terms as a function of temperature and asymmetry for $a^{\eta N}$=1.02 fm. At zero magnetic field and asymmetry,  we anticipate that the first range term gives a significant repulsive contribution to the in-medium mass whereas the mass and $d^\prime$ terms give little and significant  attractive contributions, respectively.  For the non-zero magnetic field (asymmetry), the modification in the $d^\prime$ term becomes more (less). This behavior is due to the presence of nucleon's scalar density terms  in the second term of the self-energy expression [Eq. (\ref{sek})]. 
The $d^\prime$ term dependence emphasizes the importance of scattering length $a^{\eta N}$,  in the  eta-nucleon interactions.



\begin{figure}[h]
\includegraphics[width=16cm,height=16cm]{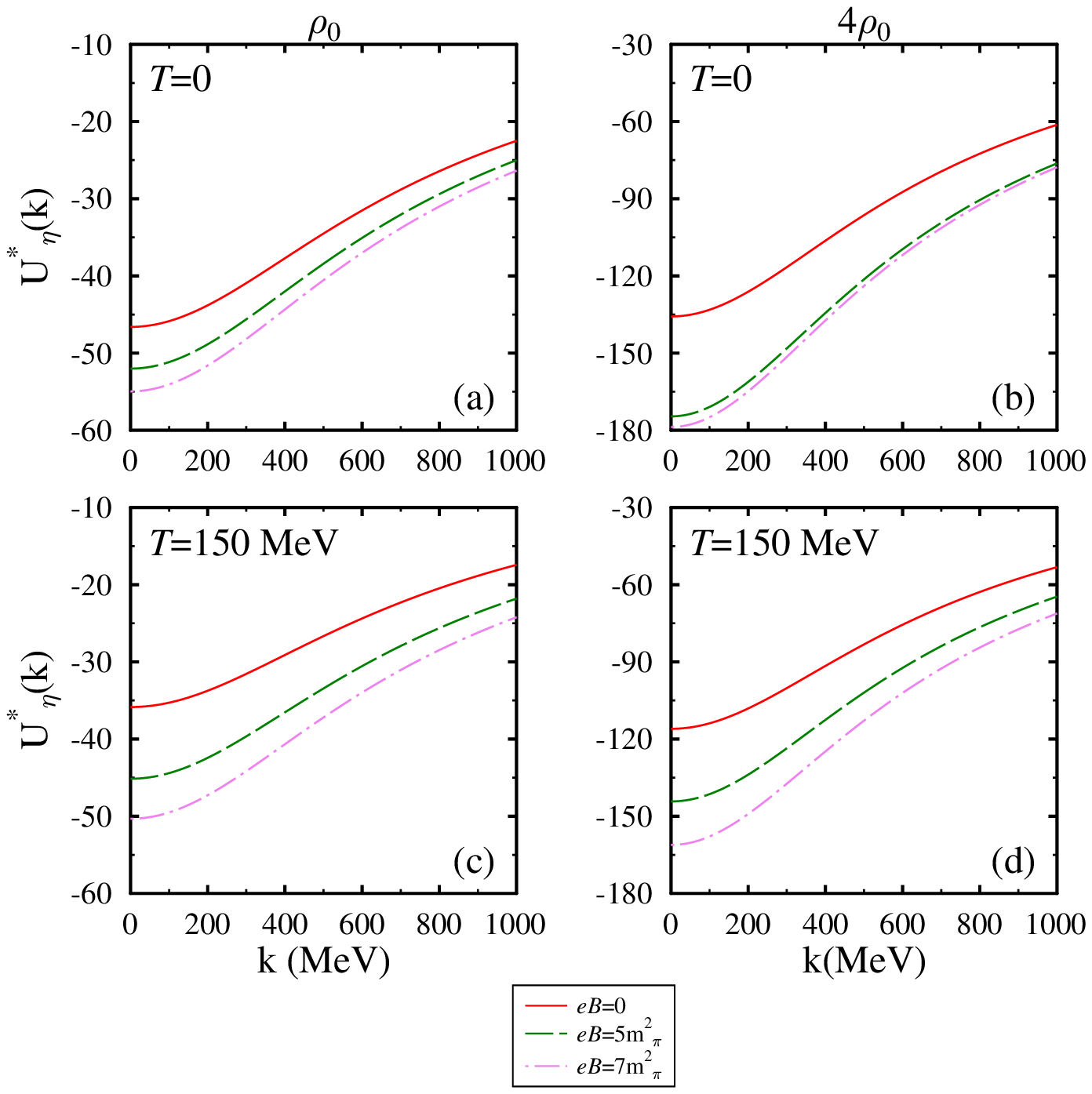}
\caption{(Color online) The in-medium $\eta$ meson optical potential in chiral model  at $a^{\eta N}$=0.91 fm and $I$=0.}
\label{f6}
\end{figure}

\begin{figure}[h]
\includegraphics[width=16cm,height=16cm]{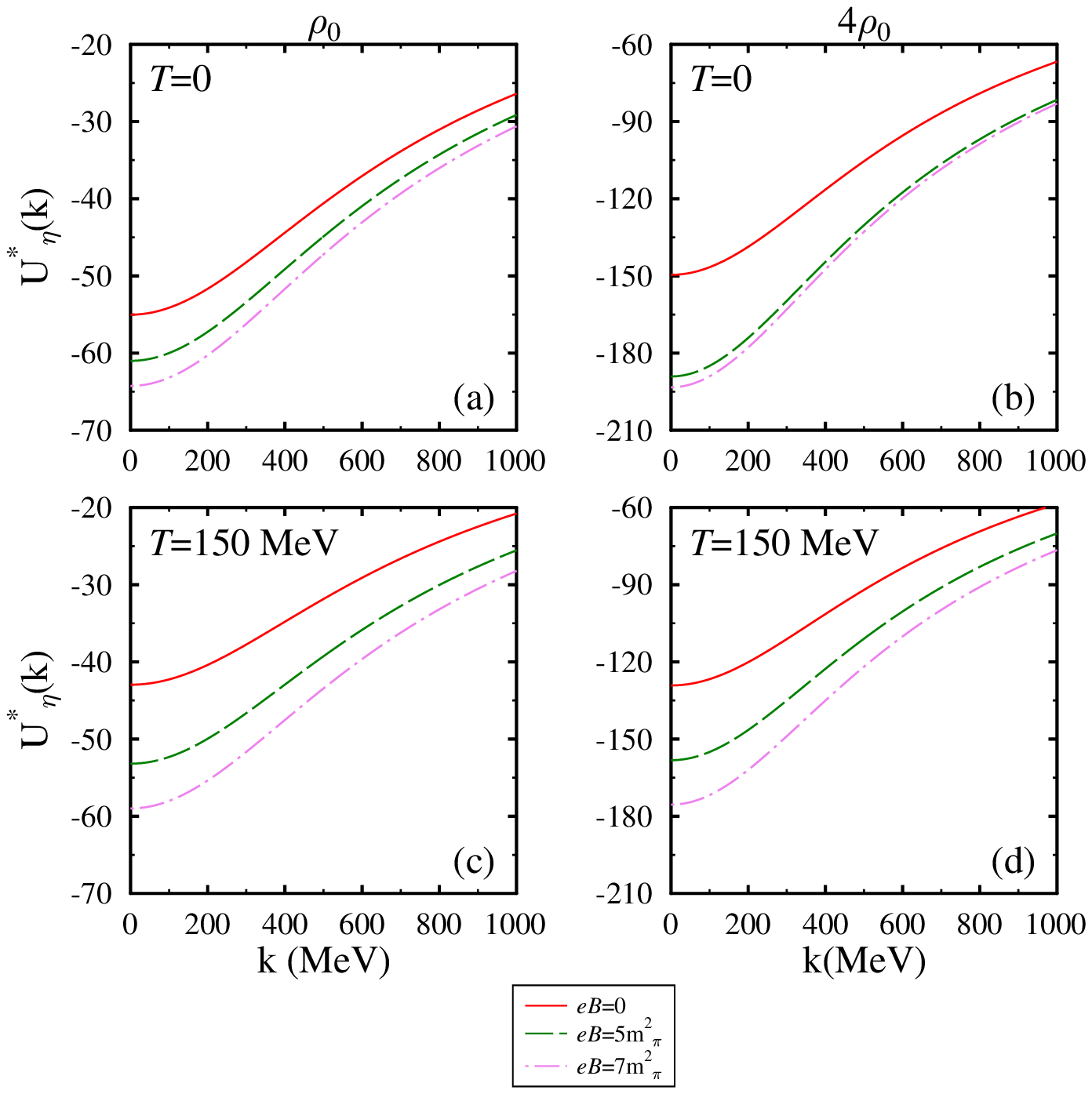}
\caption{(Color online) The in-medium $\eta$ meson optical potential in chiral model  at $a^{\eta N}$=1.02 fm and $I$=0.}
\label{f7}
\end{figure}

\begin{figure}[h]
\includegraphics[width=16cm,height=16cm]{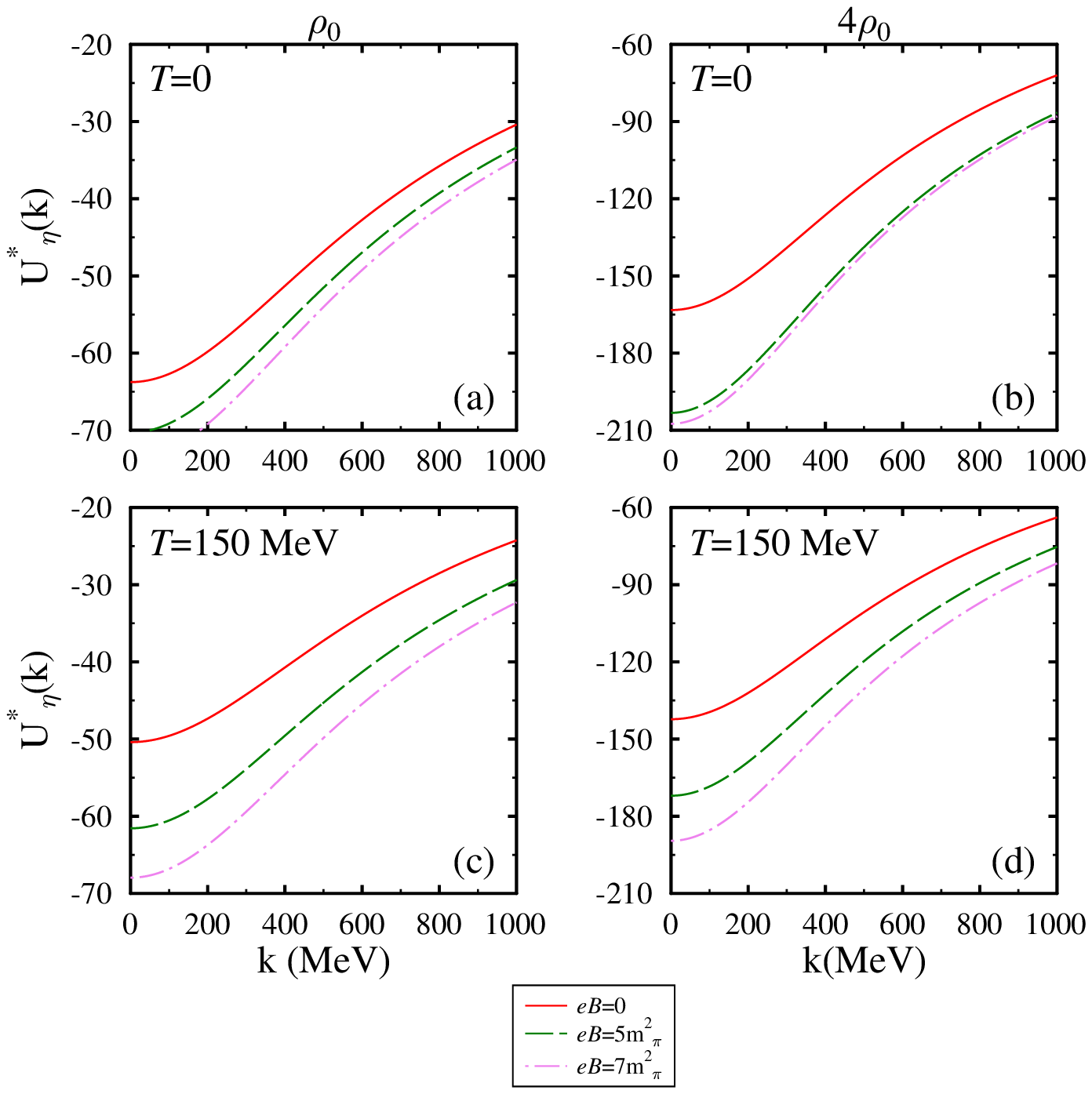}
\caption{(Color online) The in-medium $\eta$ meson optical potential in chiral model  at $a^{\eta N}$=1.14 fm and $I$=0.}
\label{f8}
\end{figure}

The optical potential at zero and non zero momentum can be used to study the $\eta$-mesic nuclei \cite{Jenkins1991,Kumar2020c,Zhong2006} and eta-meson momentum dependence \cite{Berg1994,Chen2017,David2018}. In \Cref{f6},  in symmetric nuclear matter we plot the optical potential as a function of medium momentum $\lvert  \textbf{k}\rvert$ for various values of  magnetic field and density at $a^{\eta N}$=0.91 fm.  In this figure at $\rho_N$=$\rho_0$, we observe that the magnitude of optical potential decreases with the increase in momentum. With the increase in the magnetic field (temperature), we observe the drop in optical potential become more (less).  The behavior of in-medium  optical potential  reflects the interplay between the in-medium mass and momentum which can be understood from the expression given by Eq. (\ref{opk}). At higher values of the momentum  $\lvert \textbf{k} \rvert$,  in the optical potential curve,  the contribution of effective mass is  suppressed   by the increase in momentum  states. Furthermore, in the right panel ,i.e., high density regime, we anticipate deep optical potential which becomes less as momentum states increase.   In \Cref{f7,f8}, we find likewise trend of optical potential with $\eta$ momentum. In these figures, we find a more deep optical potential with the increasing scattering length. The behavior of optical potential with scattering length and other medium parameters can be understood in terms of in-medium mass.   For a more clear picture, we  listed  the values of in-medium optical potential in chiral SU(3) model at $\lvert  \textbf{k} \rvert$=0   in \Cref{tablems}.

		 \begin{table}
\begin{tabular}{|c|c|c|c|c|c|c|c|c|c|}
\hline
& & \multicolumn{4}{c|}{$I$=0}    & \multicolumn{4}{c|}{$I$=0.5}   \\
\cline{3-10}
&$a^{\eta N} (\text{fm})$ & \multicolumn{2}{c|}{T=0} & \multicolumn{2}{c|}{T=100 }& \multicolumn{2}{c|}{T=0}& \multicolumn{2}{c|}{T=100 }\\
\cline{3-10}
&  &$eB$=0&$eB$=5${{m^2_{\pi}}}$ &$eB$=0 &$eB$=5${{m^2_{\pi}}}$ & $eB$=0 &$eB$=5${{m^2_{\pi}}}$&$eB$=0&$eB$=5${{m^2_{\pi}}}$\\ \hline 
& 0.91&-46.60& -52& -38.24 &-46.36&-45.17&-42.78&-38.21&-38.23 \\ \cline{2-10}

$ \Delta m^*_\eta$&1.02&-55&-61  &-45.66  & -54.59 & -53.42 &-50.69&-45.62 & -45.64\\ \cline{2-10}

&1.14&-63.75&-70.30 &-53.41&-63.14 & -62 &-59& -53.36 &-53.38\\ \cline{1-10}

\end{tabular}
\caption{In-medium mass-shift (MeV)  of $\eta$-meson with and without taking the effect of magnetic field at $\rho=\rho_0$  and different parameters calculated in the chiral SU(3) model.}
\label{tablems}
\end{table}

	\subsection{Impact of Magnetic Field on the $\eta$ Mesons in ChPT+Chiral Model}
	\label{subsec:3.2}

	In this section, we evaluate the in-medium mass of $\eta$-meson mass using the joint approach of chiral SU(3) model and chiral perturbation theory and also compared it with the results calculated in the chiral SU(3) model alone.  As discussed in the \Cref{subsec2.1.2}, the $\eta N$ equation of motion is derived from the ChPT  $\eta N$ Lagrangian density. The magnetic field influenced  scalar density of nucleons in the ChPT self-energy [Eq. (\ref{sekchpt})]  is taken from the chiral SU(3) model discussed in \Cref{subsec2.1}. In the present work, we took the value of parameter $\Sigma_{\eta N}$ as 280 MeV by neglecting the uncertainties in the parameter \cite{Kumar2020c}. We will see later that the contribution of  $\Sigma_{\eta N}$ term is very less as compared to the kappa term.

\begin{figure}[h]
\includegraphics[width=16cm,height=16cm]{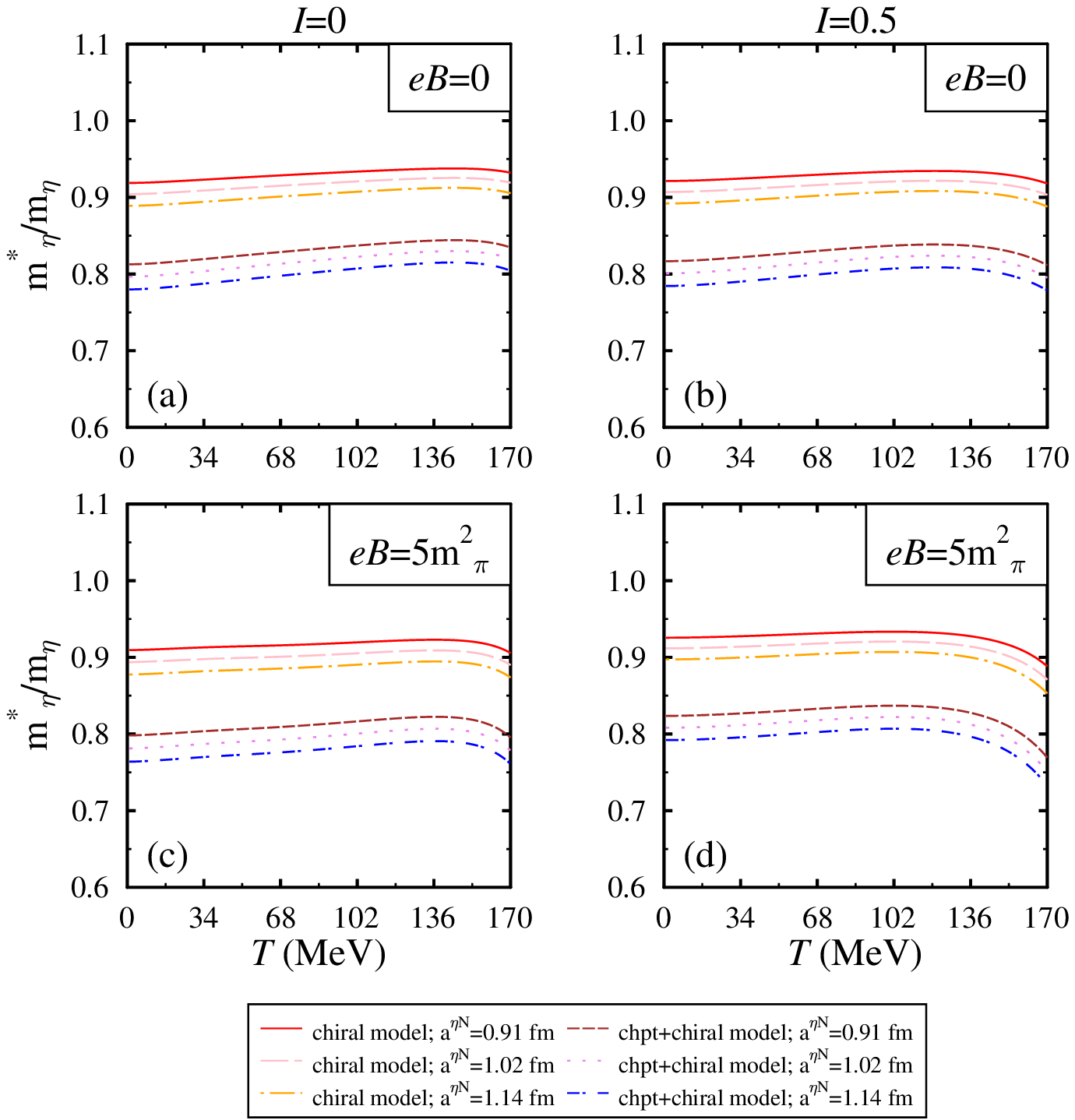}
\caption{(Color online) Comparison of  in-medium $\eta$ meson mass  at $\rho_0$. }
\label{f9}
\end{figure}

\begin{figure}[h]
\includegraphics[width=16cm,height=16cm]{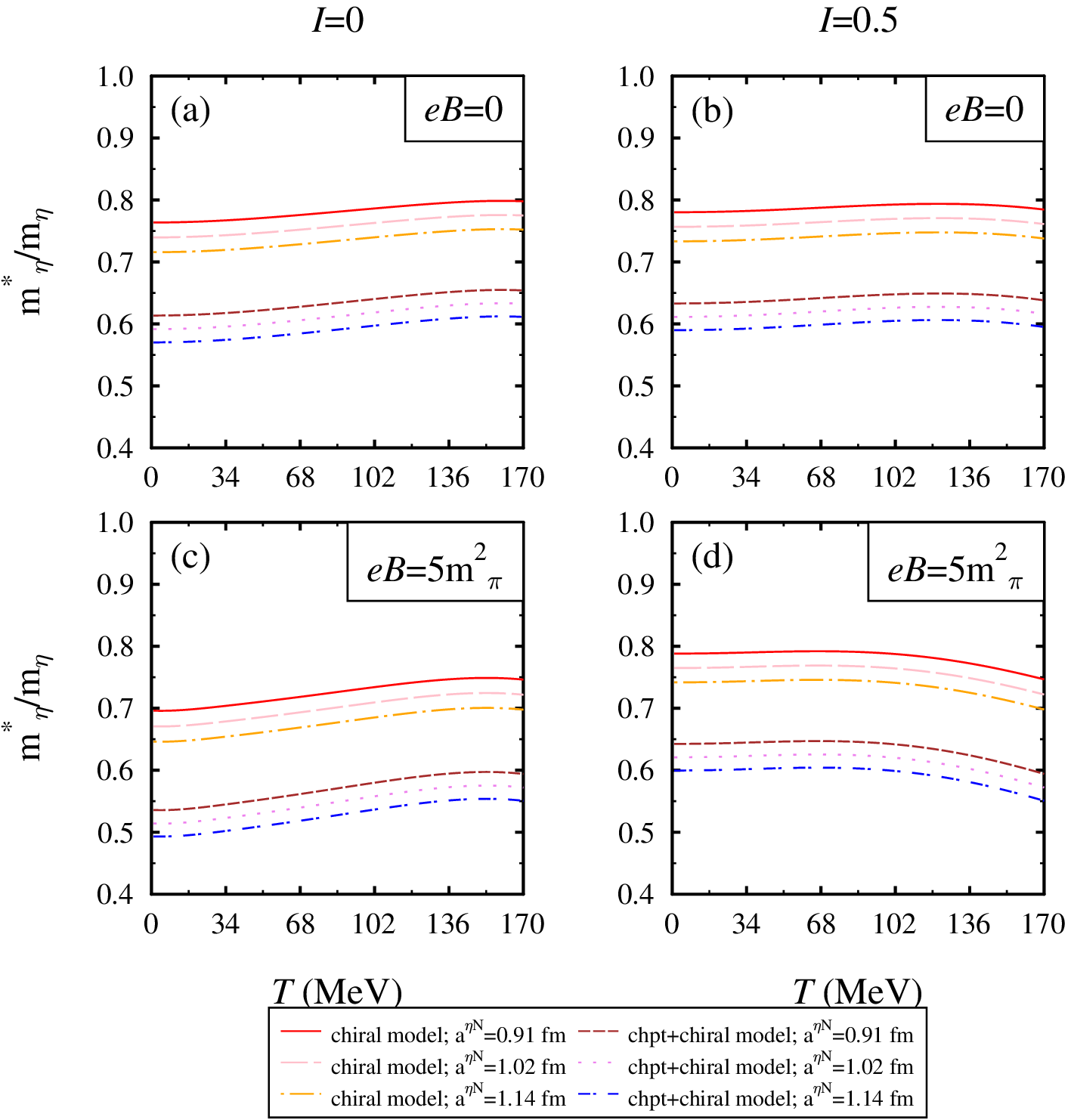}
\caption{(Color online) Comparison of  in-medium $\eta$ meson mass  at 4$\rho_0$. }
\label{f10}
\end{figure}

\begin{figure}[h]
\includegraphics[width=16cm,height=16cm]{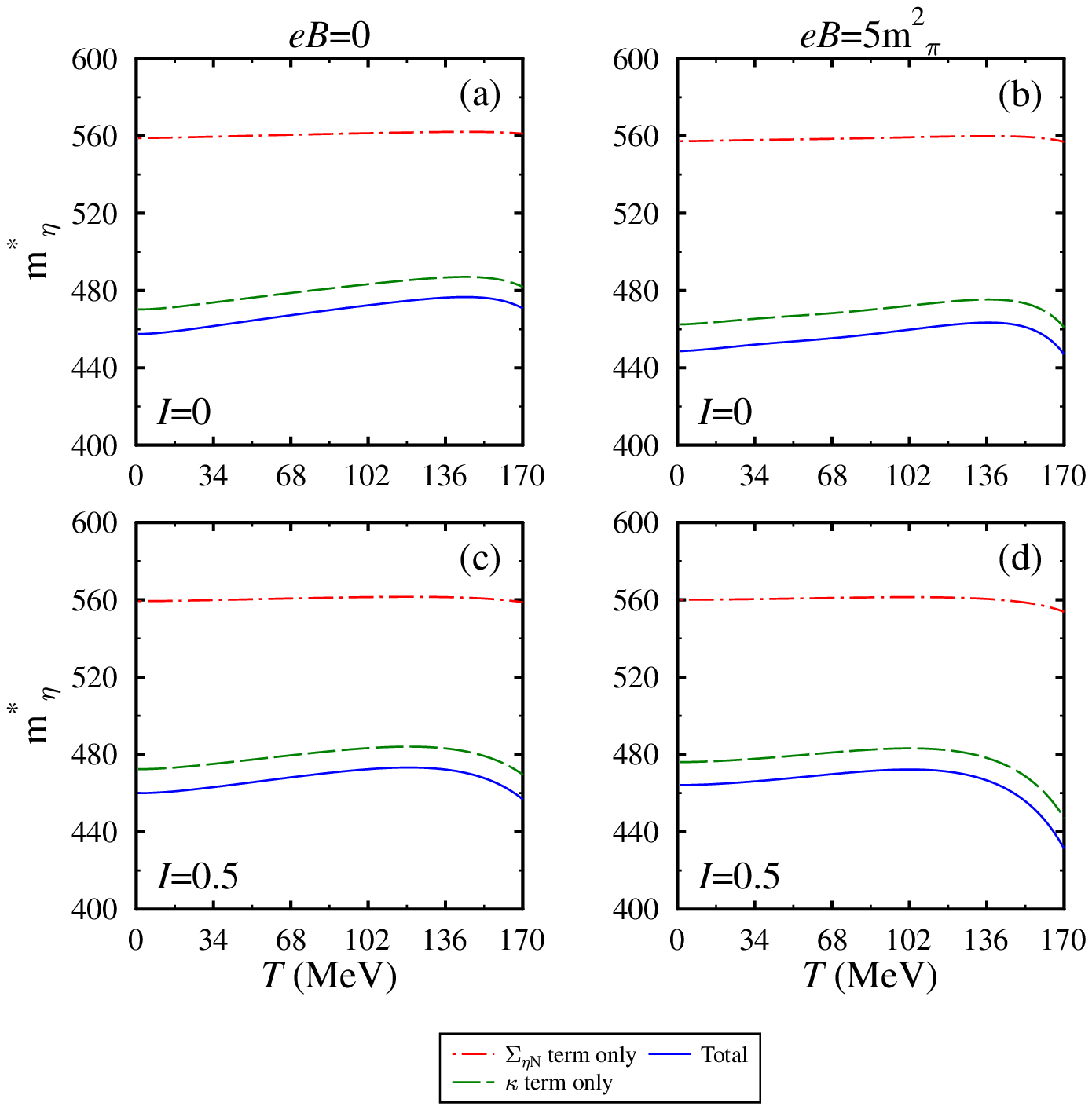}
\caption{(Color online) The different terms of  in-medium $\eta$ meson mass in ChPT+chiral model at $\rho_0$ and $a^{\eta N}$=1.02 fm. }
\label{f11}
\end{figure}

	In \Cref{f9,f10}, we plot the mass ratio $m^*_\eta/m_\eta$  with respect to temperature, scattering length and isospin asymmetry at $\rho_N$=$\rho_0$ and 4$\rho_0$, respectively. In these figures, we have also compared the in-medium mass evaluated from the  different approaches i.e. (i) ChPT and chiral model (ii) chiral model alone. In \Cref{f9}, at nuclear saturation density, using the joint approach,  we observe a significant decrement in the in-medium mass of $\eta$-meson. We find a similar behavior of the medium modified $\eta$-meson mass concerning the magnetic field, isospin asymmetry, and scattering length as was found in the observations of the chiral SU(3) model. The substantial decrement in the joint approach lies in the fact that  there is no term having repulsive contribution term in the ChPT. The net contribution in ChPT comes from the $\Sigma_{\eta N}$ and $\kappa$ term (both attractive in nature). In \Cref{f10}, at a higher value of nuclear density, we observe that the trend of mass ratio with temperature remains the same but here we get more negative mass-shift. The ratio shows similar behavior concerning other medium parameters and scattering length. To have a clear understanding, 	in \cref{f11} at $\rho_N$=$\rho_0$ and $a^{\eta N}$=1.02 fm, we have  illustrated the in-medium behavior of  individual terms present in the ChPT self-energy in magnetized asymmetric nuclear matter which contribute to  the in-medium mass of $\eta$-meson through Eq. (\ref{metac}). From this figure, we observe that the contribution of $\Sigma_{\eta N}$ term is negative but very less as compared to $\kappa$ term. The $\kappa$ term has a significant attractive contribution to the in-medium mass because, in the  in-medium mass mathematical relation given by Eq. (\ref{metac}), the numerator has negative contribution of nucleon scalar density whereas the denominator  has a  positive contribution. Therefore, due to this inverse relationship, with the increase in scalar density  the value of effective mass decreases. The values of in-medium optical potential at zero momentum  calculated using ChPT+chiral model  are given in  \cref{tablems1}.
	
	To the best of our knowledge, no work has been done to study the effect of magnetic field on the in-medium mass of $\eta$-mesons. The current results at zero magnetic field can be compared with the existing literature \cite{Zhong2006,Waas1997,Tsushima1998,Song2008}. In our previous work at zero magnetic field, we have elaborately compared the results and observed that the findings of different papers are in agreement for varying values of scattering length \cite{Kumar2020c}.

			 \begin{table}
\begin{tabular}{|c|c|c|c|c|c|c|c|c|c|}
\hline
& & \multicolumn{4}{c|}{$I$=0}    & \multicolumn{4}{c|}{$I$=0.5}   \\
\cline{3-10}
&$a^{\eta N} (\text{fm})$ & \multicolumn{2}{c|}{T=0} & \multicolumn{2}{c|}{T=100 }& \multicolumn{2}{c|}{T=0}& \multicolumn{2}{c|}{T=100 }\\
\cline{3-10}
&  &$eB$=0&$eB$=5${{m^2_{\pi}}}$ &$eB$=0 &$eB$=5${{m^2_{\pi}}}$ & $eB$=0 &$eB$=5${{m^2_{\pi}}}$&$eB$=0&$eB$=5${{m^2_{\pi}}}$\\ \hline 
& 0.91&-107.57& -115.95& -93.79 &-105.70&-105.25&-102.11&-93.72&-93.70 \\ \cline{2-10}

$ \Delta m^*_\eta$&1.02&-116.83&-125.69 &-102.21  & -114.89 & -114.35 &-110.22&-102.11 & -102.17\\ \cline{2-10}

&1.14&-126.36&-135.64 &-110.96&-124.32 & -123.75 &-119.42& -110.86 &-110.93\\ \cline{1-10}

\end{tabular}
\caption{In-medium mass-shift (MeV)  of $\eta$-meson with and without taking the effect of magnetic field at $\rho=\rho_0$  and different parameters calculated in the ChPT+chiral SU(3) model.}
\label{tablems1}
\end{table}

\section{SUMMARY}
\label{sec:4}

To summarize, we studied the effect of an external magnetic field on the 	 in-medium mass of $\eta$-meson in the hot asymmetric nuclear matter. We studied the in-medium $\eta N$ interactions using two separate methodologies. In the first approach,  we computed the in-medium mass-shift of $\eta$-meson using the chiral SU(3) model and observed a decrement in the effective mass as a function of the magnetic field and nuclear density. We anticipated substantial   medium effects  in the regime of  high magnetic field and density. In the second  approach, we used the combined method of chiral perturbation theory (ChPT) and chiral SU(3) model to compute the in-medium properties of $\eta$-meson. In the latter approach, we introduced  the medium effects through the  nucleon scalar density which is calculated in the  chiral SU(3) model. Using the joint approach, we found a substantial decrease  in the mass of $\eta$-meson concerning the magnetic field and nuclear density which is much deeper than the observations of the first approach.  The effects of  isospin asymmetry and temperature  are also incorporated and found to be a little repulsive. In both approaches, we observe a direct dependence of negative mass-shift with $a^{\eta N}$ scattering length. Furthermore, due to zero charge on the $\eta$ meson, we do not observe Landau quantization therefore no additional energy levels were discovered. The optical potential at finite momentum can be used to study the experimental properties such as     momentum dependence  \cite{David2018,Chen2017,Berg1994} and $\eta$-meson production rate \cite{Peng1987,Martinez1999,Agakishiev2013} in the magnetized nuclear medium. Also, the observed negative mass-shift can be used  to study  the possibility of $\eta N$ bound states formation \cite{Jenkins1991,Zhong2006}. Also, the magnetic field influenced optical potential may be used in future experiments to study the in-medium observables of $\eta$-mesons \cite{Rapp2010,Vogt2007}.

\section*{Acknowledgment}

One of the authors (R.K.) sincerely acknowledges the 
support of this work from Ministry of Science and Human
Resources Development (MHRD), Government of India, via 
the National Institute of Technology Jalandhar.

\end{document}